\documentclass[10pt, journal]{IEEEtran}
\IEEEoverridecommandlockouts

\usepackage[utf8]{inputenc} %
\usepackage[T1]{fontenc}    %
\usepackage[table,xcdraw]{xcolor}
\usepackage{pifont}
\usepackage[colorlinks,
    linkcolor=red,citecolor=citecolor,urlcolor=blue]{hyperref}       %
\usepackage{booktabs}       %

\usepackage{algorithm}
\usepackage{graphicx}
\usepackage{todonotes}

\usepackage{amssymb}
\usepackage{threeparttable}
\usepackage{multirow}
\usepackage{amsmath}
\usepackage{bm}
\usepackage{float}
\usepackage{amsthm}
\usepackage{soul}
\usepackage[noend]{algpseudocode}
\usepackage{enumitem} %
\usepackage[subrefformat=parens,labelformat=parens]{subfig}

\usepackage{xspace}

\usepackage{array}

\usepackage{xcolor}

\definecolor{citecolor}{RGB}{34,139,34}
\definecolor{mydarkblue}{rgb}{0,0.08,1}
\definecolor{mydarkgreen}{rgb}{0.02,0.6,0.02}
\definecolor{mydarkred}{rgb}{0.8,0.02,0.02}
\definecolor{mydarkorange}{rgb}{0.40,0.2,0.02}
\definecolor{mypurple}{RGB}{111,0,255}
\definecolor{myred}{rgb}{1.0,0.0,0.0}
\definecolor{mygold}{rgb}{0.75,0.6,0.12}
\definecolor{myblue}{rgb}{0,0.2,0.8}
\definecolor{mydarkgray}{rgb}{0.,0.2,0.2}

\definecolor{lightred}{RGB}{255,235,235}
\definecolor{lightgreen}{RGB}{235,255,235}
\definecolor{lightblue}{RGB}{235,235,255}
\definecolor{lightcyan}{RGB}{235,255,255}
\definecolor{lightmagenta}{RGB}{255,235,255}
\definecolor{lightyellow}{RGB}{255,255,235}

\definecolor{qxkcolor}{RGB}{215,235,255}
\definecolor{softmaxcolor}{RGB}{230,235,255}
\definecolor{probxvcolor}{RGB}{255,255,235}

\definecolor{topkcolor}{RGB}{255,235,235}
\definecolor{zecolor}{RGB}{255,255,235}
\definecolor{dynacolor}{RGB}{235,255,255}

\definecolor{reviewcolor}{RGB}{0,0,200}

\newcommand{\ceil}[1]{\lceil #1 \rceil}

\theoremstyle{plain}

\theoremstyle{definition}

\newtheorem{myproblem}{\textbf{Problem}}

\algdef{SE}[DOWHILE]{Do}{doWhile}{\algorithmicdo}[1]{\algorithmicwhile\ #1}%

\newcommand{\squishlist}{
 \begin{list}{$\bullet$}
  { \setlength{\itemsep}{0pt}
     \setlength{\parsep}{3pt}
     \setlength{\topsep}{3pt}
     \setlength{\partopsep}{0pt}
     \setlength{\leftmargin}{1.5em}
     \setlength{\labelwidth}{1em}
     \setlength{\labelsep}{0.5em} } }
     
\newcommand{\squishend}{
  \end{list}  }

\newcommand{\name}{\texttt{LiDAR}\xspace}

\graphicspath{{./figs/}}

\AtBeginDocument{%
  }

\usepackage{pifont}
\usepackage[normalem]{ulem}

\algdef{SE}[DOWHILE]{Do}{doWhile}{\algorithmicdo}[1]{\algorithmicwhile\ #1}%

\begin{document}
\makeatletter
\newcommand*\mytitlefontsize{\fontsize{23}{15.5}\selectfont}
\makeatother

\title{
LiDAR 2.0: Hierarchical Curvy Waveguide Detailed Routing for Large-Scale Photonic Integrated Circuits
}

\author
{
Hongjian Zhou,
Haoyu Yang,~\IEEEmembership{Member,~IEEE},
Ziang Yin,
Nicholas Gangi,
Zhaoran (Rena) Huang,~\IEEEmembership{Senior Member,~IEEE},
Haoxing Ren,~\IEEEmembership{Fellow,~IEEE},
Joaquin Matres,
Jiaqi Gu,~\IEEEmembership{Member,~IEEE}
\thanks{The preliminary version has been accepted by the IEEE/ACM International Symposium on Physical Design (ISPD) in 2025.}
\thanks{H.~Zhou and J.~Gu are with the School of Electrical, Computer and Energy Engineering, Arizona State University, AZ, USA (e-mail: jiaqigu@asu.edu).
N.~Gangi and Z.~Huang are with the Department of Electrical, Computer, and System Engineering, Rensselaer Polytechnic Institute, NY, USA.
H.~Yang and H.~Ren are with NVIDIA, TX, USA.
J.~Matres is with GDSFactory, CA, USA.}
}

\thispagestyle{empty}

\IEEEaftertitletext{\vspace{-20pt}}
\maketitle
\bstctlcite{IEEEexample:BSTcontrol} 
\begin{abstract}
\label{abstract}
Driven by innovations in photonic computing and interconnects, photonic integrated circuit (PIC) designs advance and grow in complexity. Traditional manual physical design processes have become increasingly cumbersome.
Available PIC layout tools are mostly schematic-driven, which has not alleviated the burden of manual waveguide planning and layout drawing.
Previous research in PIC automated routing is largely adapted from electronic design, focusing on high-level planning and overlooking photonic-specific constraints such as curvy waveguides, bending, and port alignment. As a result, they fail to scale and cannot generate DRV-free layouts, highlighting the need for dedicated electronic-photonic design automation tools to streamline PIC physical design.
In this work, we present \name, the first automated PIC detailed router for large-scale designs. It features a grid-based, curvy-aware A$^\ast$ engine with adaptive crossing insertion, congestion-aware net ordering, and insertion-loss optimization. To enable routing in more compact and complex designs, we further extend our router to hierarchical routing as \name~2.0. It introduces redundant-bend elimination, crossing space preservation, and routing order refinement for improved conflict resilience.
We also develop and open-source a YAML-based PIC intermediate representation and diverse benchmarks, including TeMPO, GWOR, and Bennes, which feature hierarchical structures and high crossing densities.
Evaluations across various benchmarks show that \name~2.0 produces nearly DRV-free layouts, achieving up to 16\% lower insertion loss and 7.69× speedup over prior methods on spacious cases, and 9\% lower insertion loss with 6.95× speedup over LiDAR 1.0 on compact cases.
Our codes are open-sourced at \href{https://github.com/ScopeX-ASU/LiDAR}{link}.
\end{abstract}

\vspace{-10pt}
\section{Introduction}
\label{sec:Introduction}

\begin{figure}
    \centering
    \includegraphics[width=0.95\columnwidth]{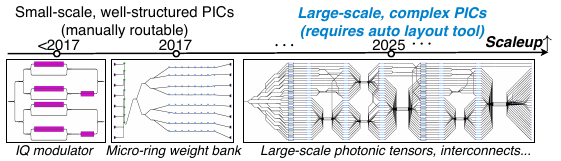}
    \vspace{-5pt}
    \caption{Modern PIC scale and complexity require EPDA.}
    \label{fig:layout}
     \vspace{-15pt}
\end{figure}

In recent years, photonic integrated circuits have attracted considerable attention due to their advantages in high-speed data transmission and low power consumption. Notable advances include photonic tensor cores (PTCs)~\cite{NP_Nature2025_Hua} developed for optical neural networks (ONNs)~\cite{NP_Nature2025_Ahmed}, as well as photonic network-on-chips (NoCs)~\cite{NP_ICCAD2022_Taheri} designed for high-bandwidth on-chip communication. Driven by these emerging applications, PIC designs are rapidly increasing in complexity. As shown in Fig.~\ref{fig:layout}, the number of photonic components per chip is approaching the scale of 1000 and is expected to double approximately every 2.5 years~\cite{thylen2006moore}. This growth highlights the urgent need for advanced electronic-photonic design automation (EPDA) toolchains~\cite{NP_ICCAD2019_Pond} to streamline layout, boost productivity, and ensure quality.

Traditionally, the physical design of PICs follows a schematic-driven approach~\cite{korthorst2023photonic}, where components are placed and connected based on the circuit topology and signal path defined in the schematic. This method seeks to minimize waveguide crossings, detours, and bends to reduce insertion loss and enhance signal integrity. In certain structured designs, routing can even be \textbf{manually managed}, especially when the circuit features a \emph{well-organized, no-crossing topology}, as seen in crossbar arrays~\cite{NP_Nature2021_Feldmann}, triangular or rectangular meshes~\cite{NP_NATURE2017_Shen}, or binary tree structures~\cite{hashemi2022review}. When such designs are \emph{optimally placed with ample spacing and well-aligned ports}, devices can be directly abutted or connected via simple straight waveguides, resembling the standard cell-based layout methodology commonly used in SRAM array design.

However, as PIC designs grow in complexity, significant routing challenges emerge, making automated routing increasingly essential. These challenges typically arise when:
\ding{202}~the \textbf{circuit scale surpasses the limits of manual design}, involving hundreds or thousands of components and nets;
\ding{203}~the circuit features a \textbf{complex topology or suboptimal placement}, resulting in port misalignments, excessive crossings, and severe routing congestion;
\ding{204}~the design must be \textbf{adaptable across different fabrication technologies or device variants}, each introducing variations in component size and behavior that impact waveguide routing;
\ding{205}~\textbf{frequent design iterations} are required, making manual updates time-consuming and error-prone, especially as schematic designers often lack full visibility into waveguide routing constraints or the necessity for inserting crossings.
Such scenarios frequently lead to repeated back-and-forth between schematic and layout teams, particularly during layout exploration phases involving iterative placement and routing refinement.

Most existing studies primarily focus on \underline{global routing planning} for PICs. A number of optical routing algorithms have been proposed for on-chip 3D system-on-package designs~\cite{minz2007optical, ding2009oil}, with the primary goal of optimizing signal loss and overall power consumption. Tools such as PROTON~\cite{proton} and PLATON~\cite{von2016platon} offer automatic placement and routing, employing a modified Lee’s algorithm for waveguide routing. In~\cite{chuang2018planaronoc}, further improvements are achieved by optimizing device orientation, such as flipping and rotation, to reduce insertion loss through fewer crossings.
While these global routing techniques are effective in identifying low-loss paths, they often \textbf{focus solely on logical path planning} and tend to \textbf{neglect physical realizability}. As a result, they may suffer from routing congestion, infeasible bend or crossing insertions, ultimately leading to an invalid routing solution.

Some research efforts have also targeted the \underline{detailed routing} stage in PIC layout design. 
The work~\cite{condrat2012methodology} proposed a method using mixed integer programming for global routing, followed by Manhattan grid-based detailed routing with crossings as constraints. 
However, this grid-based method only supports 90$^\circ$ bends. 
A follow-up approach~\cite{condrat2013channel} introduced non-Manhattan channel routing to better accommodate curved waveguides. Despite this, modern PICs typically rely on a single optical waveguide layer, making \emph{crossings unavoidable} and leaving limited flexibility for crossing optimization. 
To address these challenges, a fully automated PIC router is needed that is \textbf{physically aware of waveguide and component instantiations} and capable of performing \textbf{design-rule-compliant routing with intelligent crossing insertion}.

In this work, we propose \name2.0, a hierarchical PIC detailed routing tool that adopts a bottom-up routing strategy, supports non-Manhattan curvy waveguides, and enables adaptive crossing insertion.
It addresses key limitations of existing methods by jointly optimizing insertion loss and enforcing layout constraints, with full awareness of waveguide, bend, and crossing geometries.
Unlike schematic-driven approaches that require manual crossing insertion, \name2.0 adaptively inserts crossings during routing and produces nearly design-rule-violation (DRV)-free layouts within minutes, significantly reducing the need for post-routing fixes or iterative schematic updates.
The main contributions are summarized as follows.
\begin{enumerate}
\item We propose a fully automated PIC detailed routing tool, \name2.0, that generates real \texttt{GDSII} layouts with low insertion loss, supporting curvy waveguide geometries and automatic crossing insertion for large-scale photonic circuits within minutes.

\item \textbf{PIC Routing Benchmark}: We introduce a hierarchical PIC intermediate representation (PIC IR) and open-source scalable benchmark generators for PICs, spanning diverse routing complexities, enabling realistic and challenging evaluation of PIC routing algorithms.

\item \textbf{Curvy-Aware Non-Manhattan A$^\ast$ Routing}: We develop a customized A$^\ast$ search algorithm with adaptive neighbors to efficiently handle curvy structures and non-Manhattan routing patterns.

\item \textbf{Conflict-Resilient Routing}: We boost routability through accessibility-enhanced port assignment, congestion and crossing space penalties, and group-based net ordering with congestion-penalized rip-up and reroute, effectively reducing access conflicts and unnecessary crossings.

\item Evaluations show that \name2.0 supports PIC routing with various crossing sizes and bend radii, achieving DRV-free layouts with 16\% lower insertion loss (IL) and 7.69$\times$ speedup over prior methods on spacious benchmarks, and nearly DRV-free layouts with 9\% lower IL and 6.95$\times$ speedup over \name1.0 on compact benchmarks.

\end{enumerate}

\vspace{-8pt}
\section{Preliminaries}
\label{sec:Preliminaries}
\vspace{-3pt}
This section first reviews related VLSI routing methods and outlines key differences from PIC routing by examining PIC-specific design rules. We then describe the conventional manual PIC routing flow, followed by routing evaluation metrics and associated challenges. Notations used in this paper are summarized in Table~\ref{tab:notation}.
\begin{table}
\centering
\caption{Notations used in this paper.
}
\vspace{-5pt}
\resizebox{8cm}{!}{
\begin{tabular}{|c|p{6.8cm}|}
\hline
Symbol & Description \\ \hline
$N$                & The set of nets specified in the circuit netlist. \\
$n_i$              & The $i^\mathrm{th}$ net in $N$, $1 \leq i \leq |N|$. \\
$P$                & The set of all paths. \\
$p_i$              & The $i^\mathrm{th}$ path in $P$, $1 \leq i \leq |P|$. \\
$IL(p_i)$          & The insertion loss of the $p_i$. \\
$IL_{max}$         & The maximum insertion loss over all paths. \\
$IL_{wg}(p_i)$     & The propagation loss of the path. \\
$IL_{cr}(p_i)$     & The crossing loss of the path. \\
$IL_{bn}(p_i)$     & The bending loss of the path. \\
$\alpha_w$, $\alpha_c$, $\alpha_b$        & Coefficient of $IL_{wg}(p_i)$, $IL_{cr}(p_i)$, and $IL_{bn}(p_i)$. \\
$g_i$              & The $i^\mathrm{th}$ port group. \\
$w_{g_i}$  & Check region of group-based congestion penalty. \\
$w_{cr}$   & Check region of crossing space penalty. \\
$\lambda_c$        & The coefficient of congestion and crossing space penalty. \\
$s$        & Routing grid size. \\
\hline
\end{tabular}
}
\label{tab:notation}
\vspace{-10pt}
\end{table}

\vspace{-8pt}
\subsection{VLSI Detailed Routing}
VLSI detailed routing must address challenges such as complex design rules, pin accessibility, and limited routing resources~\cite{Routing_ISPD22_Posser}. Common routing approaches~\cite{Routing_TCAD22_Kahng, Routing_DAC12_Gester, Routing_ICCAD19_Li, Routing_ICCAD18_Sun} employ path-finding algorithms like A$^\ast$ search or maze routing, typically supported by a design rule checking (DRC) engine. A widely adopted strategy to resolve routing conflicts is \textit{negotiation-based routing}~\cite{Routing_FPGA95_McMurchie}, which iteratively applies rip-up and reroute techniques to eliminate routing failures.

One of the key differences between VLSI and PIC routing lies in the routing geometry. While VLSI routing typically follows Manhattan or unidirectional styles, PIC routing requires curvilinear waveguides. To support more flexible layouts, octagonal routing and other non-Manhattan styles have been explored in analog~\cite{Routing_Access23_Martins, Routing_TCAD14_Ou}, PCB~\cite{Routing_ICCAD10_Yan, Routing_DAC23_Liu, Routing_DAC21_Lin}, and package routing~\cite{Routing_DAC23_Chung, Routing_ICCAD23_Lee}. However, research specifically targeting curvy path routing remains limited.

\vspace{-10pt}
\subsection{Photonic Design Rules}
PIC routing typically operates on a single waveguide layer, where all nets are 2-pin optical paths. 
We summarize key design rules and highlight photonic-specific considerations.

\begin{figure}
    \vspace{-5pt}
    \centering
    \subfloat[]{\includegraphics[width=0.315\columnwidth]{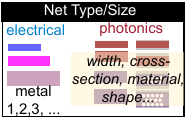}
    \label{fig:EOC_type}}
    \hfill
    \subfloat[]{\includegraphics[width=0.315\columnwidth]{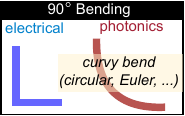}
    \label{fig:EOC_bending}}
    \hfill
    \subfloat[]{\includegraphics[width=0.315\columnwidth]{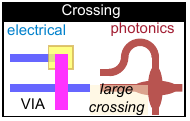}
    \label{fig:EOC_crossing}}

    \vspace{-10pt}

    \subfloat[]{\includegraphics[width=0.315\columnwidth]{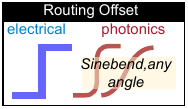}
    \label{fig:EOC_offset}}
    \hfill
    \subfloat[]{\includegraphics[width=0.315\columnwidth]{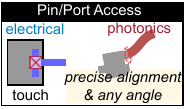}
    \label{fig:EOC_access}}
    \hfill
    \subfloat[]{\includegraphics[width=0.31\columnwidth]{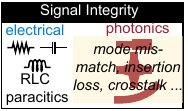}
    \label{fig:EOC_integrity}}

    \caption{Compare properties/rules of EIC and PIC routing.}
    \label{fig:EOComparison}
    \vspace{-10pt}
\end{figure}

\subsubsection{Waveguide Type}
\label{sec:wg_type}
PICs often contain multiple types of waveguides, whose characteristics depend on factors such as wavelength, polarization mode, substrate type, and cross-section, as shown in Fig.~\ref{fig:EOC_type}.
Waveguide connections must be strictly matched in type or transitioned via tapers to minimize loss. Different waveguide types also have different spacing requirements to prevent unwanted crosstalk. For high-index contrast systems (e.g., silicon-on-insulator), relatively small spacings of 1--3\,$\mu$m are typically sufficient.

\subsubsection{Bend Radius}
In contrast to the sharp 90$^\circ$ bends used in VLSI metal routing, waveguides in PICs require smooth bends to minimize mode mismatch and radiation losses.
To mitigate these losses, bends are typically implemented as circular arcs, Euler curves, or sine curves, depending on the routing scenario, as illustrated in Fig.~\ref{fig:EOC_bending}.

The bend radius can vary significantly from a few microns to millimeters, depending on material properties and waveguide design.
For example, silicon waveguides with high refractive index contrast support small bend radii of 5–10~$\mu m$, while silicon nitride waveguides, which have lower index contrast, typically require larger bend radii of 20–100~$\mu m$.
Although larger bend radii reduce insertion loss, they also increase area consumption and may limit routing flexibility.

\subsubsection{Waveguide Crossing}
Unlike VLSI routing, which forbids wire crossings and uses vias for layer transitions, PICs permit waveguide crossings (CRs) on the same layer, which are often necessary for high-density designs.
Each crossing introduces insertion loss, typically 0.1–1 dB, and occupies an area of approximately 5$\times$5 $\mu m^2$.

The crossing angle is also critical: CRs should ideally intersect at 90$^\circ$ to minimize crosstalk.
This constraint complicates routing in dense layouts, as parallel waveguides require extra space to adjust their direction via bending before forming perpendicular intersections, as shown in Fig.~\ref{fig:EOC_crossing}.

\subsubsection{Port Connection and Alignment}
Waveguides are connected via precise port abutment, which requires exact face-to-face alignment.
When there is an offset between ports, additional bending is required to compensate for the misalignment, as illustrated in Fig.~\ref{fig:EOC_offset}.
Misalignment or offset at the ports can break the optical path, making \emph{precise alignment} essential for successful routing.
An example of correct alignment is shown in Fig.~\ref{fig:EOC_access}.

\subsubsection{Signal Integrity}
Insertion loss is a critical metric for evaluating PIC routing quality, as it directly affects the laser power budget and signal integrity, including signal-to-noise ratio and crosstalk, as illustrated in Fig.~\ref{fig:EOC_integrity}
The primary evaluation metric is the maximum insertion loss along the critical path.
Long waveguides and excessive crossings degrade signal integrity and should be minimized wherever possible.

\vspace{-8pt}
\subsection{Schematic-Driven PIC Layout}
Traditional PIC layout workflows, including manual design and current EPDA tools, are schematic-driven~\cite{chrostowski2016schematic}. In this approach, all structures, including crossings and waveguide segments, are instantiated explicitly. Designers must plan routing during the schematic stage and manually insert crossings, while waveguide connections rely on port abutment without explicit physical net definitions.

A major limitation is that waveguide paths and crossings must be predetermined by design experts during schematic creation, based largely on empirical estimates of the final layout. Once these paths are defined, it becomes difficult to modify them during physical design, leading to a \textbf{rigid and manually constrained routing topology}. This inflexibility often results in repetitive revisions between the schematic and layout stages, which can be time-consuming and error-prone, and becomes impractical for large-scale designs.

To overcome these limitations, it is essential to adopt a \textbf{new formulation of instances and nets} that decouples schematic design from physical implementation. This will enable \textbf{automated crossing insertion and routing flexibility}, improving efficiency and scalability in PIC layout design.

\vspace{-8pt}
\subsection{PIC Routing Quality Metrics}
In addition to standard routing metrics such as wirelength, design rule violations, and runtime, a key photonic-specific metric is the \textbf{insertion loss (IL)}, which directly affects the link power budget and signal-to-noise ratio.
IL is calculated based on the optical path, which represents the trajectory of light propagation through all cascaded components from the laser source to the photodetector.

Assume a path $p_i$ consists of alternating instances and nets $(m_0 \rightarrow n_0 \rightarrow m_1 \rightarrow n_1 \rightarrow \cdots)$. Some instances and nets may be shared across different paths.
For simplicity and due to the lack of accurate port-specific IL data in available open-source PDKs, we assume uniform IL between any input-output port pair in multi-port photonic devices.
Nevertheless, port-specific ILs can be incorporated into the same formulation when available.
The total insertion loss $IL(p_i)$ of path $p_i$ is defined as the sum of insertion losses from all devices $IL(m_j)$ and all routed waveguide segments $IL(n_j)$ along the path. The loss is measured in decibels (dB), following standard convention.
For net-level IL, we consider three types of losses: crossing loss ($IL_{cr}$), bending loss ($IL_{bn}$), and propagation loss ($IL_{wg}$). 
Therefore, we have:
\begin{equation}
\small
\label{eq:iloss}
\begin{aligned}
    &IL(p_i) = \!\!\!\sum_{m_j\in p_i}IL(m_j)+\sum_{n_j\in p_i} IL(n_j)\\
    \sum_{n_j\in p_i}&IL(n_j)=IL_{wg}(p_i) + IL_{cr}(p_i) + IL_{bn}(p_i)\\
    IL_{wg}(p_i) = \alpha_w&  WL_{p_i}, \ 
    IL_{cr}(p_i) = \alpha_c  \# CR_{p_i}, \ 
    IL_{bn}(p_i) = \alpha_b  \angle BN_{p_i},
\end{aligned}    
\end{equation}
where $WL_{p_i}$ is the total waveguide length, $\#CR_{p_i}$ is the number of waveguide crossings, and $\angle BN_{p_i}$ is the cumulative bend angle along the path. The coefficients $\alpha_w$, $\alpha_c$, and $\alpha_b$ represent the insertion loss per unit waveguide length, per crossing, and per unit bend angle, respectively, and are determined by the specific photonic technology and component structures.
Minimizing insertion loss is crucial to achieving the desired optical performance and signal integrity for switching, modulation, or multiplexing applications.

The \textbf{maximum insertion loss} across all paths, denoted as $IL_{max}$, determines the worst-case optical budget and sets the minimum required laser power to ensure reliable detection at the outputs. Therefore, $IL_{max}$ serves as the primary evaluation metric for PIC routing. The routing objective is expressed as:
\begin{equation} \label{eq:ilmax}
IL_{max} = max_{p_i \in P}~IL(p_i)
\end{equation}

\vspace{-8pt}
\subsection{Problem Formulation} \label{subsec:problem_formulation}
We formulate the PIC detailed routing problem as follows:

\begin{myproblem}[PIC Detailed Routing]
\textit{Given a set of nets $N = {n_i \mid 1 \leq i \leq |N|}$ and a set of placed devices $M = {m_i \mid 1 \leq i \leq |M|}$, generate a routing solution for each net $n_i \in N$ such that all connections are completed without design rule violations, while minimizing $IL_{max}$.}
\end{myproblem}

\section{\name: Automated PIC Detailed Routing}
\label{sec:Method}

\begin{figure}[t]
    \centering
    \vspace{-5pt}
    \includegraphics[width=0.75\columnwidth]{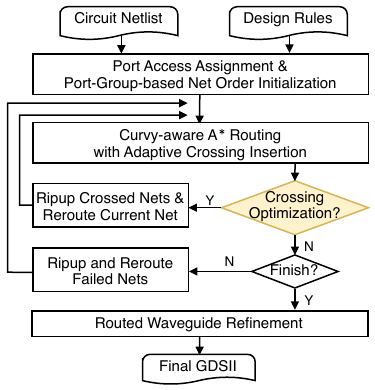}
    \vspace{-8pt}
    \caption{Algorithm flow of our \name framework.}
    \label{fig:oview}
    \vspace{-10pt}
\end{figure}

With an input circuit netlist, as shown in Fig.~\ref{fig:oview},
we present \name, a detailed PIC routing framework based on customized grid-based A$^\ast$ search. 
It efficiently generates curvy waveguide paths with automatic crossing insertion, minimizing maximum insertion loss while ensuring design-rule compliance. 
The core routing flow has three phases: 
\ding{202}~\emph{Port Access Assignment}: assigns ports by orientation and local density to ease routing and reduce congestion; 
\ding{203}~\emph{Iterative Curvy-Aware Routing}: routes all nets with a curvy-aware A$^\ast$ guided by group-based net ordering; 
\ding{204}~\emph{Crossing Optimization and Refinement}: applies local rip-up and reroute to resolve conflicts and optimize crossings, then refines results to remove redundant bends and produce a clean GDS layout.

\vspace{-8pt}
\subsection{Accessibility-Enhanced Port Assignment}
\label{sec:PortAssign}

\begin{figure}
    \centering
    \vspace{-10pt}
    \subfloat[]{\includegraphics[width=0.49\columnwidth]{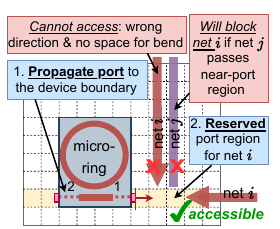}
    \label{fig:CheckOrientation}
    }
    \vspace{-10pt}
    \subfloat[]{\includegraphics[width=0.45\columnwidth]{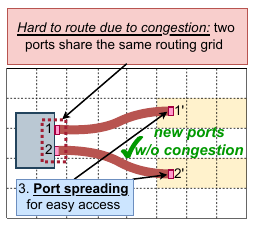}
    \label{fig:PortSpreading}
    }\\
    \subfloat[]{\includegraphics[width=\columnwidth]{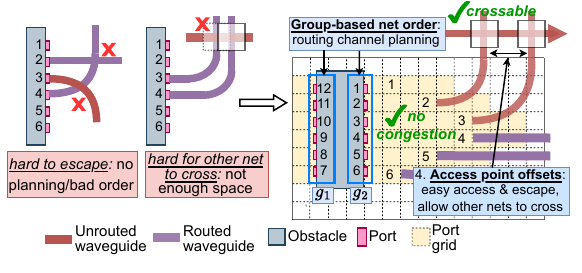}
    \label{fig:NetGroup}
    }
    \vspace{-5pt}
    \caption{(a) Port propagation and reserved port region help port access.
    (b) Port spreading removes congested ports in the same grid.
    (c) group-based net order with access point offset enables channel planning and allows potential crossing.}
    \label{fig:CheckRegion}
    \vspace{-10pt}
\end{figure}

Port accessibility is one of the most critical and challenging subroutines in PIC detailed routing.
Unlike unidirectional metal pins in VLSI, PICs use \emph{directional waveguide ports} that require exact face-to-face orientation and precise alignment, as shown in Fig.~\ref{fig:CheckOrientation}.
Improperly oriented waveguides cannot legally connect to the target port, especially when insufficient space is available to reorient the waveguide using bends.
This problem becomes even more pronounced when nearby waveguides obstruct the port region.
This difficulty stems from the large area needed for curvy bends.
To address this, we propose port access assignment techniques that consider both port orientation and spatial density to enhance accessibility.

\begin{algorithm}[t]
\caption{Port Assignment Procedure}
\label{alg:port_assignment}
\begin{algorithmic}[1]
\Require Set of components with directional ports ($\mathcal{C}$), crossing size ($s$), bend radius ($r$)
\Ensure Port assignment result and reserved regions
\For{each component $c$ in $\mathcal{C}$}
    \For{each direction $d$ in \{0, 90, 180, 270\}}
        \For{ports $p \in d$ }
            \State Sort $p$ based on coordination
            \If{$p$ lie inside $c$'s bounding box}
                \State \textbf{Port Propagation}
            \EndIf
            \If{$p$ share the same routing grid}
                \State \textbf{Congested Port Spreading}
            \EndIf
            \State Reserved\_length $\gets f(\text{index}(p), r, s)$
            \State Assign reserved region along $p$ orientation
        \EndFor
    \EndFor
\EndFor
\end{algorithmic}
\end{algorithm}

\noindent\underline{\textbf{Port Propagation}}.~
Some waveguide ports are inside the bounding box of a device. Since devices are treated as routing blockages, we propagate these internal ports outward to the device boundary based on their orientations, as illustrated in Fig.~\ref{fig:CheckOrientation}. This ensures they are reachable during routing.

\noindent\underline{\textbf{Congested Port Spreading}}.~
Certain PIC devices feature dense clusters of ports that may overlap on the same routing grid, leading to access conflicts.
To alleviate this, we apply a \emph{symmetric spreading strategy}, where high-density ports are redistributed with a predefined spacing and extension length, as shown in Fig.~\ref{fig:PortSpreading}.
These new access ports are connected to their original locations via sine bends. We add 5 units of extension length for each grid shift, ensuring they occupy distinct routing tracks and satisfying the minimum bend radius.

\noindent\underline{\textbf{Staggered Port Access Region Reservation}}.~
To prevent waveguides from blocking ports, a region in front of each port is reserved along its orientation (Fig.~\ref{fig:CheckOrientation}). 
For multiport devices such as MMIs, assigning the same reserved length to all ports can still cause blockage, while closely spaced parallel waveguides hinder crossing insertion (Fig.~\ref{fig:NetGroup}), since a crossing requires a minimum footprint and cannot be placed consecutively without sufficient spacing. 
To address this, we introduce the \emph{staggered port access region} strategy, ensuring adjacent access points are spaced larger than the crossing footprint, leaving sufficient room for accessibility. 
The reserved length depends on the port’s order among ports of the same orientation, as well as crossing size and bend radius. 
Ports are grouped by orientation and sorted by coordinates along the perpendicular axis, each assigned an index $p_i$. 
The reserved length for port $p_i$ is calculated by:
\begin{equation} \label{eq:plength}
(\frac{p_{num}}{2} - \left|\, p_i - \frac{p_{num}}{2}\,\right|) \times s +r
\end{equation}
where $p_{num}$ is the number of ports in the group. 
This yields a staggered, mountain-shaped profile that prevents inner ports from being blocked and facilitates escape. 
If routing fails, the reserved length is adjusted according to the updated routing order in Section~\ref{lab:conflict_optimization}.

\vspace{-8pt}
\subsection{Port-Group-based Net Order}
\label{sec:interOrder}
\name is a sequential router that processes one net at a time, where the routing order has a significant impact on both quality and feasibility.
To address this, we propose a port-group-based net ordering strategy that clusters ports on the same device by direction, denoted as $g_i$, e.g., 0$^\circ$ and 180$^\circ$ ports form groups $g_1$ and $g_2$ in Fig.~\ref{fig:NetGroup}
The routing proceeds \textbf{group by group based on an inter-group routing order}. Within each group, nets are routed in a defined \textbf{intra-group order} before moving to the next group.
This strategy is motivated by the observation that \textbf{most congestion and routing conflicts occur between nets in the same group}.
By routing nets group-wise with mutual awareness, intra-group conflicts are effectively reduced, improving overall routing quality.

The inter-group routing order of  $g_i$ is first determined by its group priority score $S_{g_i}=\text{min}_{n_k\in g_i}dist_{n_k}$, where $dist_{n_k}$ is the Euclidean distance of net $n_k$. A smaller $S_{g_i}$ indicates a higher routing priority for that group. If $S_{g_1} = S_{g_2}$, the group that enters the priority queue earlier will be routed first. After that, we define the net routing order within the same group.

Since ports within a group form a staggered access point region, they are more accessible and are designated as target ports, while the others serve as sources to reduce conflicts. However, the access order (i.e., intra-group routing order) is critical; an improper access sequence may lead to ports being blocked by waveguides or other port regions. Therefore, we determine the access sequence based on the source ports spatial distribution. Details are provided in Section~\ref{sec:Extension}.

\vspace{-8pt}
\subsection{Non-Manhattan Waveguide Routing with Curvy-Aware A$^\ast$}
\label{ref:astar}
Unlike Manhattan-style VLSI routing, PICs require non-Manhattan paths to support smooth curves that reduce bending and insertion loss.
We propose an iterative waveguide routing algorithm based on A$^\ast$ search, supporting both 45$^\circ$ and 90$^\circ$ turns with adaptive crossing insertion. Built on a flexible VLSI-proven framework, the algorithm enables efficient multi-objective optimization for PICs.

\subsubsection{Spacing-Ensured A$^\ast$ Routing Grid Size Setting}
In \name, the routing grid size $s$ is set to be larger than the waveguide width. Since waveguides in PICs are generally wider than their ports, this setting improves routing efficiency and simplifies port access by reducing local congestion near narrow access points.

\subsubsection{Parametric Curvy-Aware Neighbor Candidate Generation}

\begin{figure}
    \centering
    \includegraphics[width=0.93\columnwidth]{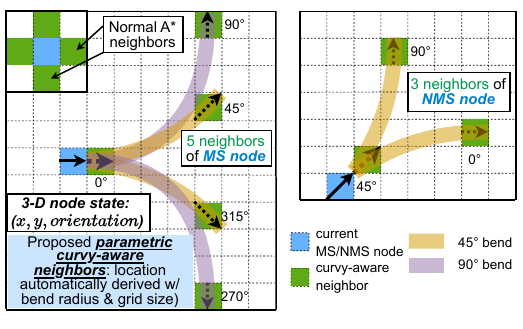}
    \vspace{-5pt}
    \caption{Parametric curvy-aware neighbors allow non-Manhattan curvy waveguide routing.
    Neighbors are automatically derived based on bending radius and grid size.
    }
    \label{fig:NeighborDefinition}
    \vspace{-10pt}
\end{figure}

To enable curvy-aware A$^\ast$ search, we propose a parametric neighbor generation scheme that leverages bending geometry and performs comprehensive design rule checks to filter out illegal candidates.

While traditional VLSI and PCB routing often use sharp 90$^\circ$ or fixed 45$^\circ$ turns, PIC routing requires smooth curves. We define each routing node by its spatial location and orientation, represented as a directional node $(x, y, \textit{orientation})$. This directional representation is essential for maintaining alignment and orientation during port access.
As illustrated in Fig.~\ref{fig:NeighborDefinition}, neighbor candidates are determined by the node's current orientation and a user-defined bending radius $r$. Based on orientation, nodes are categorized into two states: Manhattan State (MS) and Non-Manhattan State (NMS). MS nodes, which align with the x- or y-axis, have five possible neighbors: one directly forward and four at angular offsets of $\pm$45$^\circ$ and $\pm$90$^\circ$. In contrast, NMS nodes, which follow diagonal trajectories, have three potential neighbors.

The exact positions of neighbor candidates are adaptively computed using the bend radius $r$ and grid size $s$. Larger bending radii and smaller grids lead to larger step sizes. For example, for an MS node oriented at 0$^\circ$, its directly adjacent neighbor is 1 grid unit away, while the steps to reach the 90$^\circ$ and 45$^\circ$ neighbors are calculated as:
\vspace{-5pt}
\begin{equation} \label{eq:gcost}
\begin{aligned}
    & step_{90,x} = step_{90,y} = \ceil{r / s},  \\ 
    step_{45,x} = \ceil{(\sqrt{2}&-1) r/s};~ step_{45,y} = \ceil{(1-\frac{\sqrt{2}}{2}) r/s}.
\end{aligned}
\end{equation}
Here, $step_{90,x}$ and $step_{90,y}$ represent the horizontal and vertical grid distances required to reach a neighbor at a 90$^\circ$ turn.
The ceiling function $\lceil \cdot \rceil$ guarantees that the bend radius constraint is met.
Note that unlike 45$^\circ$ diagonal neighbors in traditional 8-way A$^\ast$ search, where $step_{45,x}$ is equal to $step_{45,y}$ , our $45^\circ$ neighbors are the endpoint of the 45-degree curves, where $step_{45,x}$ and $step_{45,y}$ are not equal.

\subsubsection{Geometry-Aware Neighbor Legality Check}

To ensure only feasible neighbor candidates are explored during A$^\ast$ search, each candidate must pass a geometry-aware legality check before being added to the priority queue. A neighbor is considered legal only if the corresponding waveguide segment does not violate any design rules upon geometric instantiation.

\noindent\underline{\textbf{Hit No Obstacle: Geometry-Aware Spacing Check}}.~
If the neighbor does not intersect any obstacle, we instantiate the actual geometry of the connecting waveguide segment and perform a spacing check to verify that the path complies with design rules and is free of violations (DRVs).
\begin{figure}
    \centering
    \includegraphics[width=\columnwidth]{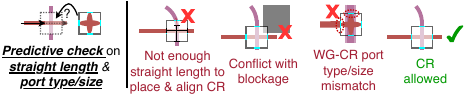}
    \vspace{-13pt}
    \caption{Proposed adaptive waveguide crossing insertion.}
    \label{fig:CrossingInsertion}
    \vspace{-12pt}
\end{figure}

\noindent\underline{\textbf{Hit Routed Nets: Predictive Crossing Insertion}}.~
If the neighbor candidate intersects with an already routed waveguide (marked as an obstacle), we evaluate whether a waveguide crossing can be inserted to legally pass through it.

As shown in Fig.~\ref{fig:CrossingInsertion}, several critical conditions must be satisfied to allow crossing insertion:
\ding{202}~\emph{Sufficient straight waveguide length}: Crossings occupy a defined footprint and require perpendicular waveguide orientations. We ensure there is enough straight length and validate the orientation by checking the routing grid state at the potential crossing location.
\ding{203}~\emph{No overlap with blockages}: The bounding box of the crossing must not intersect any existing obstacles to comply with the design rules.
\ding{204}~\emph{Port matching}: For a valid crossing, the intersecting waveguides must align with the four ports of the crossing structure. This includes matching properties such as waveguide width, cross-section, and layer.
By predictively checking all those legality conditions, our method adaptively enables crossing insertion when needed. This significantly \textbf{reduces unnecessary detours and eliminates the rigidity of manually defined crossings} in schematic-driven flows.

\subsubsection{Insertion Loss-Aware A$^\ast$ Search Cost}

\begin{figure}[t]
    \centering
    \vspace{-5pt}
    \includegraphics[width=0.8\columnwidth]{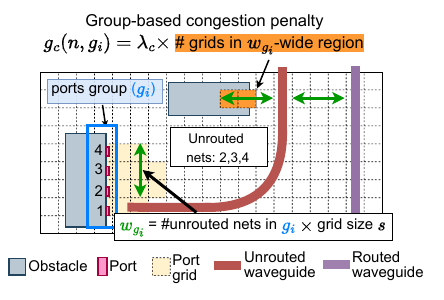}
    \vspace{-7pt}
    \caption{Left: group-based congestion penalty in Eq.~\eqref{eq:gccost}.\ Right: crossing-space penalty allows consecutive crossing insertion.}
    \label{fig:gcp}
    \vspace{-10pt}
\end{figure}

\name adopts a customized A$^\ast$ search cost function that incorporates insertion loss considerations to improve both routing quality and algorithm efficiency.
The total cost of a node $n$ in A$^\ast$ search is given by the standard formulation:
$f(n) = g(n) + h(n)$,
where $g(n)$ denotes the accumulated cost from the source node $s$ to the current node $n$, and $h(n)$ is the heuristic estimate of the cost from node $n$ to the target node $t$.
The cost $g(n)$ includes an insertion-loss-inspired term $g_{IL}(n)$, adapted from Eq.~\eqref{eq:iloss} with a tunable weight and a \textbf{group-based congestion penalty (GCP)}, denoted as $g_c(n, g_i)$ to save routing resources for a group of nets. By adjusting the weight parameters, users can flexibly balance between waveguide length and the number of crossings to meet specific design objectives:
\begin{equation} 
\small
\label{eq:gccost}
\begin{aligned}
g(n) = g_{IL}(n) + g_c(n, g_i),~~g_c(n, g_i) = \lambda_c\#grids(w_{g_i}), \\ 
\end{aligned}
\end{equation}
where $\lambda_c$ is a penalty coefficient that discourages routing too close to blockages or previously routed waveguides, preserving routing resources for unrouted nets within the same routing group, and $\#grids(w_{g_i})$ is the number of grids that occupied by others in the check region $w_{g_i}$ as shown in Fig.~\ref{fig:gcp}.
The check region $w_{g_i}$ is determined by the number of unrouted nets in its port group. 
As more nets are routed, $w_{g_i}$ decreases to avoid consuming extra space.

Since our method allows non-Manhattan routing, the shortest path is often diagonal. We define a customized heuristic $h(n)$ as shown in Eq.~\eqref{eq:hcost}.
\vspace{-6pt}
\begin{equation} 
\small
\label{eq:hcost}
\begin{aligned}
    & d_{min} = \text{min}(|n_x-t_x|, |n_y-t_y|), \\
    & d_{max} = \text{max}(|n_x-t_x|, |n_y-t_y|), \\
    & h(n) = d_{max} - d_{min} + \sqrt{2} \ast d_{min} + \alpha \cdot IL_{bd, 45}, \\
    & \alpha =
        \begin{cases} 
        1,  & \text{if } d_{min}>0 ~\text{and}~d_{max}>0 \\
        0,  & \text{others}
        \end{cases},
\end{aligned}
\end{equation}
where $d_{min}$ and $d_{max}$ represent the minimum and maximum differences in the x and y coordinates between the current node $n$ and the target node $t$, respectively. 
When $\alpha = 0$, we have $d_{min}=0$, and $h(n)$ reduces to the standard Manhattan case.
When $\alpha \neq 0$, it indicates that the subsequent path search must contain at least one bend with an angle no smaller than $45^\circ$. 
Therefore, we incorporate the loss of a $45^\circ$ bend into $h(n)$, which \textbf{does not overestimate the cost and thus the A$^*$ router remains admissible}, and penalizes the bend and encourages paths that end in orientations suited for port connection.

\subsubsection{Waveguide Instantiation}

\begin{figure}[t]
    \centering
     \vspace{-5pt}
    \includegraphics[width=0.75\columnwidth]{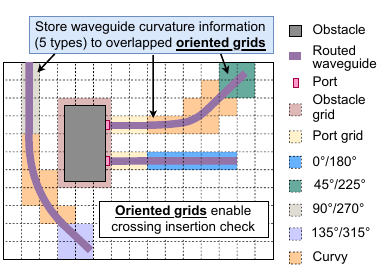}
    \vspace{-5pt}
    \caption{Represent routed waveguides in oriented grid map.}
    \label{fig:wg}
    \vspace{-12pt}
\end{figure}

A key distinction of \name compared to prior global routing methods is its \textbf{geometry awareness}.
After determining a routing path, we instantiate the actual curvy waveguide geometry using the extrude function from GDSFactory~\cite{GDSFactory}, and record it on an \textbf{overlapped, orientation-aware routing grid map}, as illustrated in Fig.~\ref{fig:wg}.
This enables the A$^\ast$ search engine to treat existing waveguides as obstacles and efficiently perform waveguide spacing checks and determine valid locations for crossing insertion.

\subsubsection{Violated Net Removal}
When the router fails to access the target port with the correct orientation, \name applies a rip-up-and-reroute (RR) strategy.
DRC is temporarily relaxed to allow routing progress, and any nets that violate design rules or conflict with existing paths are recorded. These violating nets are then removed and rerouted in subsequent iterations.
To prevent repeated routing failures and reduce congestion, a history cost~\cite{liu2013nctu} is maintained and updated in a global history map before each net is ripped up.
This history-based negotiation mechanism discourages routing over previously congested regions and helps distribute paths more evenly across the layout.
In practice, this approach effectively resolves routing failures and improves overall success by balancing routing demands across all nets.

\vspace{-8pt}
\subsection{Crossing-Waveguide Optimization}
\begin{figure}[t]
    \centering
    \includegraphics[width=0.9\columnwidth]{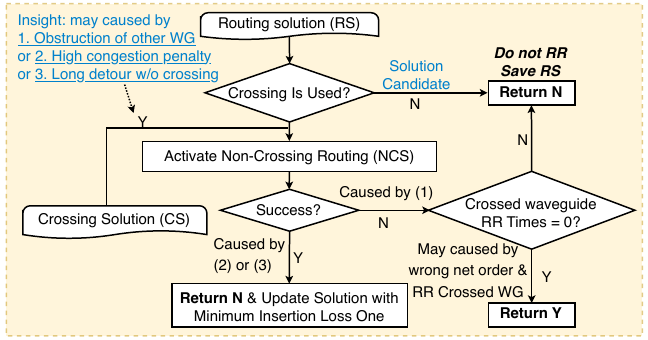}
    \vspace{-5pt}
    \caption{LRR check after finding a routing solution.}
    \label{fig:lrr}
    \vspace{-10pt}
\end{figure}

Waveguide crossings are sometimes inserted to bypass congested regions, circumvent obstructions, or avoid excessive detours. 
However, they not only introduce insertion loss, phase shifts, and occupy chip area, but they also cause crosstalk and potentially impact subsequent net routing.
Consequently, we propose a local rip-up-and-reroute (LRR) strategy to eliminate unnecessary crossings and optimize the overall routing.

As shown in Fig.~\ref{fig:lrr}, LRR is triggered after a solution is found.
If the initial routing solution (RS) contains no crossings, it is accepted directly.
If crossings exist, possible causes include: (1) a blockage caused by another waveguide requiring a crossing, (2) a crossing chosen to bypass congestion regions, or (3) high propagation loss for non-crossing paths.
To verify these, we rerun the routing with crossings disabled (NCS).
If NCS succeeds, its result is compared against RS by insertion loss, and the lower-loss path is selected.
If NCS fails, it indicates the net is blocked. 
We then check the blocking waveguide. If the blocking waveguide has never been ripped up before, it will be ripped up, as this operation will not affect CR re-insertions in subsequent iterations. 
This LRR process empirically optimizes the routing by \textbf{balancing long waveguides and CRs} to reduce overall loss.

\vspace{-8pt}
\subsection{Routed Waveguide Refinement}
\begin{figure}[t]
    \centering
    \includegraphics[width=0.75\columnwidth]{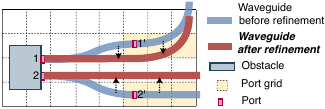}
    \caption{Waveguide refinement to remove bending.}
    \label{fig:adjustment}
    \vspace{-13pt}
\end{figure}

Since the grid-based routing may cause slight misalignment between the port center and the grid center, a small offset can occur between the routed path and the port, as illustrated in Fig.~\ref{fig:adjustment}.
To resolve this, we adjust the initial and final segments of the waveguide to precisely align with the port, without altering the bend radius along the path.
If direct adjustment is not feasible, a sine bend is introduced at the port to ensure proper alignment.

\section{Extended Conflict-Resilient Hierarchical PIC Routing Scheme}
To enable a more compact layout and ensure successful routing under limited routing resources, it is essential not only to minimize bending and crossing but also to reserve sufficient space for them, as they are area-consuming. 
This becomes even more challenging in compact layouts. To address this, we further extend our framework to \name~2.0, enhancing its conflict resolution capability and enabling successful routing even under tight layout constraints.

\label{sec:Extension}

\vspace{-8pt}
\subsection{Hierarchical Netlist Tree Construction}
\label{sec:hier}
\begin{figure}[t]
    \centering
    \includegraphics[width=0.85\columnwidth]{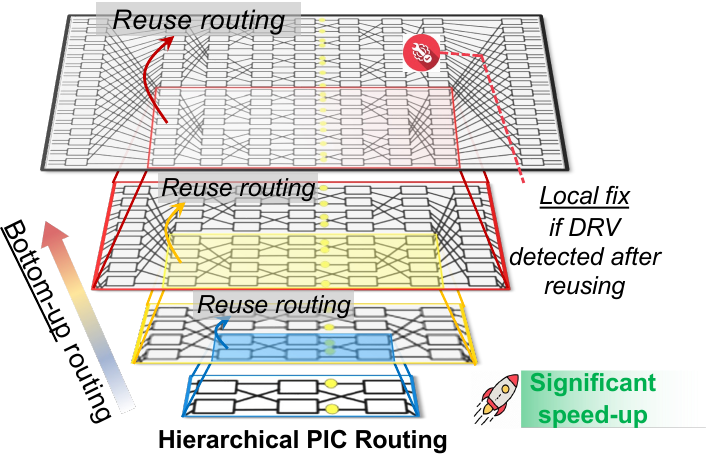}
    \caption{Hierarchical PIC netlist: different colors indicate different module types. Routing results for identical modules can be reused to improve efficiency. We perform a local fix if reusing causes minor DRV errors.}
    \label{fig:hier_circuit}
    \vspace{-10pt}
\end{figure}

PIC netlists often exhibit inherent \textbf{hierarchical structure}, where complex systems are composed of repeated functional modules, as shown in Fig.~\ref{fig:hier_circuit}. Leveraging this structure is crucial for scalable physical design. 
To exploit such hierarchy, we construct a \textit{Hierarchical Tree} by parsing the nested module instantiations defined in the YAML-format netlist, which is similar in style to a SPICE netlist and inherently preserves hierarchical relationships. 
In this tree, the \textbf{root node} corresponds to the top-level module of the circuit, the \textbf{leaf nodes} represent primitive photonic components, and the \textbf{internal nodes} represent intermediate-level submodules. The resulting tree has arbitrary fanout and depth, depending on the circuit topology. 
The construction has linear time complexity of $\mathcal{O}(n)$, where $n$ is the total number of unique modules and components. This hierarchical representation enables global routing planning at multiple levels.

We then perform \textbf{bottom-up} routing for each level of modules progressively. 
Within each level, the priority of each routing group is determined according to the group priority score defined in Subsection~\ref{sec:interOrder}.
This multi-level routing approach helps avoid conflicts between different modules, as routing is localized within each submodule before higher-level integration. 
Similar to standard-cell design in VLSI, the internal routing resources of a module are treated as private and should not be accessed by external nets.

To further improve efficiency, we \textbf{reuse routing results} of identical module types via translation, rotation, and flipping. Before reuse, we apply design rule checking at the target location to ensure legality. If a reused result violates constraints, we fall back to re-routing that specific instance from scratch. This reuse mechanism significantly accelerates layout generation for large-scale, hierarchical PIC designs.

\vspace{-8pt}
\subsection{Routing Conflict Reduction via Spatial-Aware Optimization}
\label{lab:conflict_optimization}
\begin{figure}[t]
    \centering
    \vspace{-5pt}
    \subfloat[]{\includegraphics[width=0.4\columnwidth]{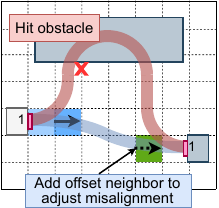}
    \label{fig:offset_neighbor}
    }
    \vspace{-5pt}
    \subfloat[]{\includegraphics[width=0.4\columnwidth]{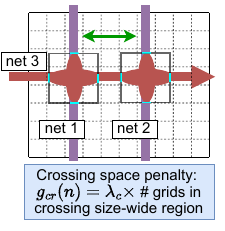}
    \label{fig:crossing_space}
    }\\

    \caption{(a) Add offset neighbor to remove redundant bending caused by port misalignment. (b) Preserve crossing space for the crossing nets.}
    \vspace{-15pt}
\end{figure}

\noindent\underline{\textbf{$A^\star$ Offset Neighbor}}.~
Although the $A^\ast$ algorithm discourages excessive bends via its bend-cost heuristic, it may still add unnecessary bends when ports are slightly misaligned. 
In tight spaces, such misalignment can cause multiple bends, raising insertion loss and risking routing failure (Fig.~\ref{fig:offset_neighbor}). 

To address this, we add \textbf{offset neighbors} under specific conditions: 
\emph{when the current node is within four bending radii of the target and their orientations differ by 180 degrees}. 
These neighbors correct small misalignments (below the bend radius) between source and target ports. 
Their locations are computed analytically from the bend radius and grid size, ensuring smooth, fabricable sine-bend waveguides. 
From an algorithmic perspective, offset neighbors simplify port access, cut redundant detours, and improve search efficiency.

\noindent\underline{\textbf{Preserved Crossing Space}}.~
A key source of routing conflict in PICs is the limited space for inserting crossings. Unlike VLSI vias, which can be inserted independently of other nets, photonic waveguide crossings must be placed directly on top of the crossed waveguides and require significant area, making insertion difficult in dense regions.
As shown in Fig.~\ref{fig:crossing_space}, closely spaced waveguides may block required consecutive crossings, causing routing failure.

To mitigate the crossing space conflict, we introduce a \textbf{crossing space penalty} $g_{cr}(n)$ during the A$^\ast$ search to encourage the router to preserve sufficient spacing around the nets that are likely to be crossed, as inferred from the netlist topology.
The $g_{cr}(n)$ penalty is defined as:
\begin{equation}
\label{eq:crcost}
g_{cr}(n) = \lambda_c \cdot \#\text{grids}(w_{cr}),
\end{equation}
where $\lambda_c$ is a user-defined penalty and $w_{cr}$ is the minimum width required to insert a valid crossing. The penalty is applied over a $w_{cr}$-wide region around the net segment, discouraging other nets from occupying this reserved space.

\vspace{-8pt}
\subsection{Reduce Routing Conflict via Routing Order Optimization}
\label{sec:rrc}
\begin{figure}[t]
    \centering
    \vspace{-10pt}
    \subfloat[]{\includegraphics[width=0.9\columnwidth]{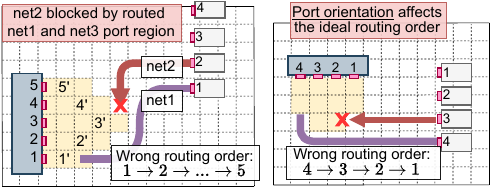}
    \label{fig:order1}
    }\\
    \vspace{-5pt}
    \subfloat[]{\includegraphics[width=0.9\columnwidth]{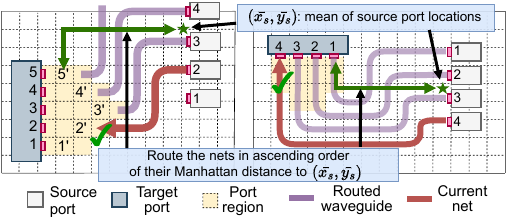}
    \label{fig:order2}
    }
    \vspace{-5pt}
    \caption{(a) Routing conflicts due to inappropriate routing order. (b) Route based on the Manhattan distance between the target port and the mean of the source port locations.}
    \label{fig:order}
    \vspace{-5pt}
\end{figure}

Although port regions allow staggered access points to reduce conflicts, an improper order may still block target ports with routed waveguides or other regions. 
For multi-port components, the \textbf{intra-group routing order} is thus crucial. 
As shown in Fig.~\ref{fig:order1}, both orientation and spatial location of ports strongly influence order, as they determine accessibility and potential conflicts.

To address this, we derive the order from the spatial relation between source and target ports (Fig.~\ref{fig:order2}). 
Let the source port set be \( \mathcal{P}_s = \{p_1, p_2, \dots, p_n\} \) with positions \( (x_i, y_i) \). 
Their mean coordinates are:
\vspace{-5pt}
\begin{equation} 
\label{eq:meanport}
(\bar{x_{s}}, \bar{y_{s}}) = \left( \tfrac{1}{n} \sum_{i=1}^{n} x_i, \tfrac{1}{n} \sum_{i=1}^{n} y_i \right)
\end{equation}
Routing order is then based on Manhattan distance from each target port to $(\bar{x_s}, \bar{y_s})$, with closer ports given higher priority. 
This method effectively resolves most conflicts, especially in multiport components.

\begin{figure}[t]
    \centering
    \vspace{-5pt}
    \includegraphics[width=0.93\columnwidth]{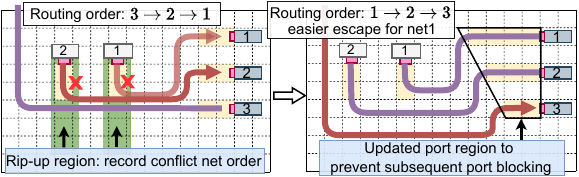}
    \caption{Rip-up blocking waveguides and update the routing order based on conflict precedence, while resizing target port regions to prevent access blockage for subsequent nets.}
    \label{fig:update}
    \vspace{-10pt}
\end{figure}

\noindent\underline{\textbf{Routing Order Refinement}}.~
It is difficult to guarantee an optimal routing order. 
When nets are routed independently without group information, the method above is inapplicable, making dynamic order adjustment necessary. 
An improper order may block ports with earlier waveguides, causing escape or access failures. 
As shown in Fig.~\ref{fig:update}, routing net1, net2, and net3 independently leads to net2’s failure, as net1’s port region and net3’s waveguide block its path.

To address routing failures, we enhance rip-up and re-route by removing not only nets directly conflicting with the failed path but also those \emph{in front of the failed net’s ports} to ease congestion.
As shown in Fig.~\ref{fig:update}, we scan along the port orientation, identify conflict order and conflicting nets (starting from \(n_0\)), and use them to adjust routing order.
For instance, if \texttt{net3} is blocked by \texttt{net2}, and \texttt{net2} by \texttt{net1}, we merge them and uniformly update the routing order from 3 to 1.
When \texttt{net3} and \texttt{net2} are both blocked by \texttt{net1} and their order is indistinguishable, we sort them by net names.
Although higher-priority nets are routed first, they may still block others. To prevent this, we dynamically \textbf{extend high-priority target ports}.
Specifically, the port length is extended by \(|N - i| \times r\), where \(N\) is the number of merged nets, \(i\) the order, and \(r\) the bend radius, shifting access points and reducing blocking risk.
Our extended conflict-resilient hierarchical routing scheme can effectively enhance routability even under tight layout constraints and help accelerate the routing process.

\vspace{-8pt}
\section{PIC Routing Benchmark}
\label{sec:benchmarks}

\begin{table}[t]
\centering
\caption{Spacious PIC benchmark statistics.}
\small
\resizebox{\columnwidth}{!}{
\begin{tabular}{|c|c|c|c|c|}
\hline
\textbf{Benchmarks} & \textbf{\# Components} & \textbf{\# Nets} & \textbf{\# Total CR} & \textbf{Die Size ($\mu m^2$)} \\
\hline
Clements\_8$\times$8     & 52  & 79  & 0   & 4800 $\times$ 1600  \\
Clements\_16$\times$16   & 168 & 287 & 0   & 8000 $\times$ 3200  \\
ADEPT\_8$\times$8        & 82  & 111 & 33  & 4400 $\times$ 1600  \\
ADEPT\_16$\times$16      & 162 & 223 & 55  & 6900 $\times$ 3200  \\
ADEPT\_32$\times$32      & 318 & 446 & 121 & 13000 $\times$ 6400 \\
Light\_(a,b,c,d)      & 9 & 16 & - & 10000 $\times$ 10000 \\
\hline
\end{tabular}
}
\vspace{-5pt}
\label{tab:bench_stat}
\end{table}

\begin{table}[t]
\centering
\caption{Compact PIC benchmark statistics.}
\small
\resizebox{\columnwidth}{!}{
\begin{tabular}{|c|c|c|c|c|}
\hline
\textbf{Benchmarks} & \textbf{\# Components} & \textbf{\# Nets} & \textbf{\# Total CR} & \textbf{Die Size ($\mu m^2$)} \\
\hline
Clements\_8$\times$8\_C     & 60    & 87    & 0    & 3500 $\times$ 1000 \\
Clements\_16$\times$16\_C   & 184   & 303   & 0    & 6800 $\times$ 2000 \\
ADEPT\_8$\times$8\_C        & 104   & 127   & 30   & 4400 $\times$ 1000 \\
ADEPT\_16$\times$16\_C      & 227   & 319   & 80   & 6200 $\times$ 2000 \\
ADEPT\_32$\times$32\_C      & 529   & 767   & 160  & 7800 $\times$ 4000 \\
TeMPO\_8$\times$8\_C           & 452   & 515   & 49   & 1700 $\times$ 1500 \\
TeMPO\_16$\times$16\_C         & 1796  & 2051  & 225  & 3000 $\times$ 3400 \\
TeMPO\_32$\times$32\_C         & 7172  & 8195  & 961  & 5700 $\times$ 6500 \\
GWOR\_16$\times$16\_C          & 17    & 32    & --   & 4000 $\times$ 4000 \\
GWOR\_32$\times$32\_C          & 33    & 64    & --   & 6000 $\times$ 6000 \\
Benes\_16$\times$16\_C        & 224   & 416   & 88   & 2200 $\times$ 1300 \\
Benes\_32$\times$32\_C        & 576   & 1024  & 416  & 3500 $\times$ 2500 \\
\hline
\end{tabular}
}
\vspace{-12pt}
\label{tab:benchmark_stats_c}
\end{table}

\noindent\underline{\textbf{PIC Intermediate Representation}}.
Inspired by the LEF/DEF formats widely used in electronic design automation, we introduce a \textbf{YAML-based photonic intermediate representation (PIC IR)} to describe the connectivity and types of photonic components. This IR supports \textbf{hierarchical module definitions}, enabling modular reuse and streamlined management of complex circuits.
To bridge this abstract representation with physical layouts, we develop a  \textbf{translation tool} that interfaces with \texttt{GDSFactory}, allowing for both visualization and generation of valid \texttt{GDSII} layouts from the IR.

\noindent\underline{\textbf{Benchmark Suites}}.
Built upon the PIC IR, we further develop and open-source a suite of \textbf{parameterized benchmark generators}. These scripts enable users to generate benchmarks with \textbf{configurable scales}, such as the number of components, number of nets, and overall circuit dimensions. They also support fine-grained control over \textbf{routing complexity} by adjusting parameters including component pitch (i.e., component density), port density, port misalignment, and the number of topological crossings, allowing for the creation of benchmarks with varying levels of routing difficulty.
The benchmark suite includes both photonic tensor core and optical switch, providing a diverse set of testcases to evaluate the router in realistic PIC settings. Users can also define their own PIC benchmarks easily by using our interface.

For the \textbf{photonic tensor core}, we provide the following benchmarks, each available at multiple scales:

\begin{itemize}
    \item \texttt{Clements}~\cite{NP_Optica2018_Clements}: A highly structured Mach-Zehnder interferometer (MZI) array with a regular topology and no topological crossings. It is relatively straightforward to route when sufficient layout space is available.
    
    \item \texttt{ADEPT}~\cite{NP_DAC2022_Gu}: An auto-searched photonic tensor core that includes numerous multiport components and a large number of inherent crossings, resulting in high routing congestion and increased difficulty.
    
    \item \texttt{TeMPO}~\cite{zhang2024tempo}: A time-multiplexed dynamic photonic tensor accelerator with a hierarchical crossbar structure. It contains multiple computing nodes and large fan-out MMI components, along with a significant number of crossings, making routing particularly challenging.
\end{itemize}

For the \textbf{optical switch} category, we use benchmarks below:

\begin{itemize}
    \item \texttt{Light}~\cite{zheng2021topro}: Based on macro-ring resonators (MRRs), it features an unstructured topology with a centralized MRR macro. We define four layout variants according to memory controller and hub placement to evaluate routing under different traffic patterns, focusing on minimizing crossings and insertion loss.
    
    \item \texttt{GWOR}~\cite{tan2011scalable}: Also MRR-based, it splits the central switch into multiple distributed parts, with I/O ports placed uniformly along the chip periphery and no group-level planning. This benchmark emphasizes routing conflict resolution via pure spatial reasoning.
    
    \item \texttt{Benes}~\cite{qiao201732}: Based on Mach-Zehnder interferometers, it adopts a structured, hierarchical Benes topology. It exhibits \textbf{extremely high crossing density}, where a single net may require tens of crossings, challenging both crossing reservation and congestion-aware routing strategies.
\end{itemize}

Tables~\ref{tab:bench_stat} and~\ref{tab:benchmark_stats_c} summarize the benchmark statistics. 
Total \#CR refers to the number of topological crossings inherent to the circuit and resolved during routing. For \texttt{Light} and \texttt{GWOR}, internal switch crossings are excluded, as we focus solely on the routing between the switch and the I/O interfaces in these benchmarks.
Table~\ref{tab:bench_stat} (ISPD version) includes spacious, moderately complex layouts, while Table~\ref{tab:benchmark_stats_c} features more \textbf{compact} benchmarks with higher component density, smaller die sizes, and more crossings, posing greater routing challenges and better reflecting real-world photonic layout constraints.
For the \texttt{Light} and \texttt{GWOR} benchmarks, the bending radius, waveguide width, and grid resolution are set to 60~$\mu$m, 2~$\mu$m, and 50~$\mu$m, respectively, to emulate the larger bend radii of the SiN platform. 
Since these circuits provide abundant routing resources, a larger grid size is adopted to accelerate the search process. 
For the other benchmarks, these values are set to 5~$\mu$m, 0.5~$\mu$m, and 2~$\mu$m, respectively, reflecting the tighter bends of the SiPh platform. By default, the crossing size is fixed at 10~$\mu$m~$\times$10$\mu$m.

\begin{table*}[t]
\centering
\caption{Comparisons of the maximum insertion loss $IL_\mathit{max}$ (dB), the path length with $IL_\mathit{max}$ (WL ($ mm$)), the number of crossings passed by the signal with $IL_\mathit{max}$, total design rule violations (\#DRV), and runtime (s) for \textbf{spacious} benchmarks. 
}
\renewcommand{\arraystretch}{1.2}
\small
\resizebox{\textwidth}{!}{
\begin{tabular}{cccccccccccccccc}
\hline \addlinespace[0.4ex]
\multirow{2}{*}{\raisebox{-0.5ex}[0pt]{Benchmarks}} & \multicolumn{5}{c}{PROTON~\cite{proton} (Adaptive crossing penalty)} & \multicolumn{5}{c}{\name 1.0~\cite{PD_ISPD2025_Zhou}} & \multicolumn{5}{c}{\name 2.0} \\
[-0.1em]
\cmidrule(lr){2-6} \cmidrule(lr){7-11} \cmidrule(lr){12-16}\addlinespace[-0.3ex]
& \#CR & WL & $IL_{\max}\downarrow$ & \#DRV $\downarrow$ & Time $\downarrow$ (s) 
& \#CR & WL & $IL_{\max}\downarrow$ & \#DRV $\downarrow$ & Time $\downarrow$ (s) 
& \#CR & WL & $IL_{\max}\downarrow$ & \#DRV $\downarrow$ & Time $\downarrow$ (s) \\
\hline
Clements\_8$\times$8    & 0  & 3.39  & 16.99 & 0  & 112 & 0  & 2.94  & 16.38 & 0  & 29  & 0  & 2.89  & \textbf{15.98} & 0  & \textbf{7} \\
Clements\_16$\times$16  & 5  & 5.06  & 29.31 & 12 & 527 & 0  & 4.38  & 26.74 & 0  & 144 & 0  & 4.07  & \textbf{26.03} & 0  & \textbf{61} \\
ADEPT\_8$\times$8       & 16 & 4.70  & \textbf{17.12} & 26 & 194 & 18 & 4.10  & 18.00 & 0  & 71  & 18 & 3.99  & 17.63 & 0  & \textbf{65} \\
ADEPT\_16$\times$16     & 28 & 7.84  & 24.07 & 98 & 1395 & 16 & 7.38  & 17.80 & 0  & \textbf{243} & 16 & 6.95  & \textbf{17.20} & 0  & \textbf{243} \\
ADEPT\_32$\times$32     & 66 & 16.13 & 44.57 & 355 & 10894 & 50 & 15.04 & 36.34 & 0  & 1348 & 50 & 14.70 & \textbf{36.06} & 0  & \textbf{1038} \\
Light\_a        & 12 & 32.98 & 11.09 & 0  & \textbf{39} & 0  & 31.11 & 7.78  & 0  & 101 & 0  & 29.99 & \textbf{7.61}  & 0  & 103 \\
Light\_b        & 6  & 18.71 & \textbf{5.89}  & 0  & \textbf{8} & 0  & 21.55 & 6.31  & 0  & 44  & 0  & 21.12 & 6.24  & 0  & 47 \\
Light\_c        & 14 & 20.81 & 10.23 & 1  & \textbf{52} & 0  & 35.29 & \textbf{8.40}  & 0  & 72  & 0  & 35.92 & 8.57  & 0  & 74 \\
Light\_d        & 13 & 28.49 & 10.94 & 1  & \textbf{53} & 0  & 33.52 & 8.14  & 0  & 80  & 0  & 32.82 & \textbf{8.09}  & 0  & 83 \\
\hline
Geo-mean        & -  & 15.35 & 18.91 & -  & 1475 & -  & 17.26 & 16.21 & -  & 237 & -  & 16.94 & \textbf{15.93} & -  & \textbf{191} \\
Ratio           & -  & 1     & 1     & -  & 1 & -  & 1.12  & 0.86  & -  & 0.16 & -  & 1.1   & \textbf{0.84}  & -  & \textbf{0.13} \\
\hline
\end{tabular}
}
\label{tab:stat}
\vspace{-10pt}
\end{table*}

\begin{table}[htp]
\centering
\caption{Device IL parameters used in $IL_{max}$ evaluation.
}
\resizebox{\columnwidth}{!}{
\begin{tabular}{|c|c|c|c|c|c|}
\hline
Propagation $\alpha_w$     &$90^\circ$ Bend $\alpha_b$ & CR $\alpha_c$ & Y-branch & MZI & MMI \\\hline
1.5 dB/cm~\cite{chan2010architectural}      & 0.005 dB~\cite{chan2010architectural} & 0.52 dB~\cite{proton} & 0.3 dB~\cite{ybrand} & 1.2 dB~\cite{mzi} & 0.1 dB~\cite{mmi} \\
\hline
\end{tabular}
}
\label{tab:parameters}
\vspace{-15pt}
\end{table}

\vspace{-8pt}
\section{Evaluation Results}
\label{sec:ExperimentalResults}

The proposed photonic detailed routing framework is implemented in Python based on GDSFactory 8~\cite{GDSFactory}
libraries. 
All evaluations are conducted on a Linux server with a 128-core AMD EPYC 7763 CPU and 512-GB memory.
Note that runtime and results may slightly differ from the ISPD version due to platform and GDSFactory version variations.

The placement solutions of all benchmark circuits are verified with simulation using GDSFactory and KLayout. Device insertion losses used for critical path IL evaluation are listed in Table~\ref{tab:parameters}, with bending loss proportional to angle.

\noindent\underline{\textbf{A$^\ast$ Search Parameters}}.~
Referring to the parameters in Table~\ref{tab:parameters} and scaling them to a unified micrometer unit, the base propagation, bending, and crossing weights are set to 1.5, 50, and 5000, respectively.
On top of these, we further adjust the weights according to the characteristics of each benchmark.
For the \textbf{photonic tensor core}, where crossings are inherent and unavoidable, we set the crossing weight equal to the bending weight to speed up search.
For the \textbf{optical switch}, the crossing weight remains 5000, while the bending weight is also set to 5000 to reduce bends; in long waveguides with large radii, shorter wirelength can offset added bend loss.
The \textbf{group congestion penalty (GCP)} parameter is determined by the detour distance, as saving routing resources through GCP may conflict with wirelength minimization. We set this value to 500, corresponding to reserving approximately 200~$\mu$m of spacing.
The \textbf{crossing spacing coefficient} is set similarly, but its checking range is limited to a single crossing.

\noindent\underline{\textbf{Baselines}}.~
We compare \name2.0 with PROTON~\cite{proton}, which performs path planning with adaptive crossing penalties but lacks valid waveguide geometry generation.
For fair comparison, we adapt PROTON with reserved port regions and global rip-up and reroute support.
If no feasible solution is found due to congestion, DRC constraints are temporarily relaxed for specific failed nets for meaningful comparison of \#DRVs.
\name1.0 is also included to demonstrate the improvements in conflict handling, crossing insertion, and runtime efficiency in \name2.0.

\vspace{-8pt}
\subsection{Results on Spacious Benchmarks}
\label{sec:SpaciousResults}

We first evaluate the routing algorithm under spacious layout conditions.
Table~\ref{tab:stat} presents a comparison of routing metrics on the spacious benchmarks.
In this setting, \name1.0 and \name2.0 achieve comparable performance and successfully generate DRV-free layouts across all benchmarks.
 Compared to PROTON, \name2.0 achieves an average of \textbf{16\% lower critical path IL and 7.69$\times$ speedup}.

\noindent\underline{\textbf{Analysis of PTC Results}}.~Despite the \texttt{Clements} circuit having a regular topology with no inherent crossings and relatively spacious layouts, insufficient planning and lack of reserved access space lead to unnecessary crossings and DRVs in PROTON. 
These routing conflicts become more severe in \texttt{ADEPT} benchmarks, which feature dense multi-port MMIs and a large number of crossings.
By incorporating \textbf{Accessibility-Enhanced Port Assignment} and \textbf{Port-Group-Based Net Ordering}, \name2.0 effectively mitigates these conflicts and reduces unnecessary rip-up and reroute.
As circuit scale increases, PROTON suffers from escalating \#DRVs and runtime, while \name2.0 consistently produces \textbf{DRV-free, low-IL layouts}.

\noindent\underline{\textbf{Analysis of Optical Switch Results}}.~
\texttt{Light} benchmark provides abundant routing resources and is primarily used to evaluate the router's ability to optimize insertion loss and crossings.
Notably, there exists a counter-intuitive trade-off between \#CR and WL; minimizing \#CR may lead to long detours that increase propagation loss, resulting in a higher overall $IL_{\max}$.
Except for the \texttt{Light\_b} case, \name achieves the lowest $IL_{\max}$ across all other cases while maintaining crossing-optimal (\#CR=0), DRV-free layouts.

\vspace{-8pt}
\subsection{Results on Compact Benchmarks}
\label{sec:CompactResults}
\begin{figure*}[t]
    \centering
    \vspace{-4pt}
    \subfloat[]{\includegraphics[width=0.7\columnwidth]{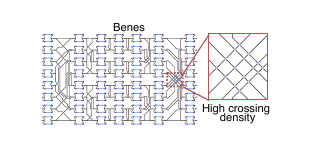}
    \label{fig:layout_a}
    }
    \vspace{-4pt}
    \subfloat[]{\includegraphics[width=0.58\columnwidth]{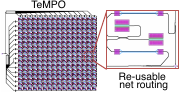}
    \label{fig:layout_b}
    }
    \vspace{-4pt}
    \subfloat[]{\includegraphics[width=0.4\columnwidth]{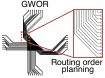}
    \label{fig:layout_c}
    }\\
    \subfloat[]{\includegraphics[width=0.93\columnwidth]{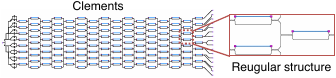}
    \label{fig:layout_d}
    }
    \subfloat[]{\includegraphics[width=0.87\columnwidth]{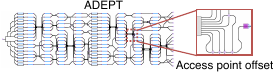}
    \label{fig:layout_e}
    }
    \vspace{-5pt}
    \caption{Layout of (a) 
    \texttt{Benes}\_16$\times$16  (b) \texttt{TeMPO}\_16$\times$16 (c) \texttt{GWOR}\_32$\times$32 
    (d) \texttt{Clements}\_16$\times$16\_C
    (e) \texttt{ADEPT}\_16$\times$16\_C generated by \name2.0.}
    \label{fig:layout_compact}
    \vspace{-5pt}
\end{figure*}

\begin{table*}[t]
\centering
\caption{Comparisons of the maximum insertion loss $IL_\mathit{max}$ (dB), the path length with $IL_\mathit{max}$ (WL ($mm$)), the number of crossings passed by the signal with $IL_\mathit{max}$, total design rule violations (\#DRV), and runtime (s) for \textbf{compact} benchmarks. 
}
\renewcommand{\arraystretch}{1.2}
\small
\resizebox{\textwidth}{!}{
\begin{tabular}{cccccccccccccccc}
\hline \addlinespace[0.4ex]
\multirow{2}{*}{\raisebox{-0.5ex}[0pt]{Benchmarks}} & \multicolumn{5}{c}{PROTON~\cite{proton} (Adaptive crossing penalty)} & \multicolumn{5}{c}{\name 1.0~\cite{PD_ISPD2025_Zhou}} & \multicolumn{5}{c}{\name 2.0} \\
[-0.1em]
\cmidrule(lr){2-6} \cmidrule(lr){7-11} \cmidrule(lr){12-16}\addlinespace[-0.3ex]
& \#CR & WL & \cellcolor[HTML]{D9D9D9}$IL_{\max}\downarrow$ & \cellcolor[HTML]{D9D9D9}\#DRV $\downarrow$ & \cellcolor[HTML]{D9D9D9}Time $\downarrow$ (s) 
& \#CR & WL & \cellcolor[HTML]{D9D9D9}$IL_{\max}\downarrow$ & \cellcolor[HTML]{D9D9D9}\#DRV $\downarrow$ & \cellcolor[HTML]{D9D9D9}Time $\downarrow$ (s) 
& \#CR & WL & \cellcolor[HTML]{D9D9D9}$IL_{\max}\downarrow$ & \cellcolor[HTML]{D9D9D9}\#DRV $\downarrow$ & \cellcolor[HTML]{D9D9D9}Time $\downarrow$ (s) \\
\hline
Clements\_8$\times$8\_C     & 0  & 2.12  & \cellcolor[HTML]{D9D9D9}16.28 & \cellcolor[HTML]{D9D9D9}2   & \cellcolor[HTML]{D9D9D9}11    & 0  & 1.91  & \cellcolor[HTML]{D9D9D9}16.23 & \cellcolor[HTML]{D9D9D9}2   & \cellcolor[HTML]{D9D9D9}11    & 0  & 1.91  & \cellcolor[HTML]{D9D9D9}\textbf{16.23} & \cellcolor[HTML]{D9D9D9}0   & \cellcolor[HTML]{D9D9D9}\textbf{9} \\
Clements\_16$\times$16\_C   & 1  & 3.91  & \cellcolor[HTML]{D9D9D9}27.13 & \cellcolor[HTML]{D9D9D9}5   & \cellcolor[HTML]{D9D9D9}93    & 0  & 3.56  & \cellcolor[HTML]{D9D9D9}26.56 & \cellcolor[HTML]{D9D9D9}5   & \cellcolor[HTML]{D9D9D9}101   & 0  & 3.56  & \cellcolor[HTML]{D9D9D9}\textbf{26.56} & \cellcolor[HTML]{D9D9D9}0   & \cellcolor[HTML]{D9D9D9}\textbf{89} \\
ADEPT\_8$\times$8\_C        & 14 & 4.53  & \cellcolor[HTML]{D9D9D9}16.26 & \cellcolor[HTML]{D9D9D9}11  & \cellcolor[HTML]{D9D9D9}96    & 14 & 3.71  & \cellcolor[HTML]{D9D9D9}16.12 & \cellcolor[HTML]{D9D9D9}2   & \cellcolor[HTML]{D9D9D9}82    & 14 & 3.71  & \cellcolor[HTML]{D9D9D9}\textbf{16.12} & \cellcolor[HTML]{D9D9D9}0   & \cellcolor[HTML]{D9D9D9}\textbf{68} \\
ADEPT\_16$\times$16\_C      & 34 & 8.24  & \cellcolor[HTML]{D9D9D9}27.60 & \cellcolor[HTML]{D9D9D9}42  & \cellcolor[HTML]{D9D9D9}515   & 21 & 8.63  & \cellcolor[HTML]{D9D9D9}21.09 & \cellcolor[HTML]{D9D9D9}7   & \cellcolor[HTML]{D9D9D9}304   & 21 & 8.35  & \cellcolor[HTML]{D9D9D9}\textbf{21.02} & \cellcolor[HTML]{D9D9D9}0   & \cellcolor[HTML]{D9D9D9}\textbf{236} \\
ADEPT\_32$\times$32\_C      & 49 & 10.78 & \cellcolor[HTML]{D9D9D9}36.20 & \cellcolor[HTML]{D9D9D9}171 & \cellcolor[HTML]{D9D9D9}1496  & 31 & 10.58 & \cellcolor[HTML]{D9D9D9}27.07 & \cellcolor[HTML]{D9D9D9}30  & \cellcolor[HTML]{D9D9D9}735   & 30 & 9.83  & \cellcolor[HTML]{D9D9D9}\textbf{26.48} & \cellcolor[HTML]{D9D9D9}3   & \cellcolor[HTML]{D9D9D9}\textbf{568} \\
TeMPO\_8$\times$8\_C           & 12 & 1.73  & \cellcolor[HTML]{D9D9D9}18.48 & \cellcolor[HTML]{D9D9D9}14   & \cellcolor[HTML]{D9D9D9}95  & 10 & 1.90  & \cellcolor[HTML]{D9D9D9}17.33 & \cellcolor[HTML]{D9D9D9}5   & \cellcolor[HTML]{D9D9D9}147   & 7  & 2.83  & \cellcolor[HTML]{D9D9D9}\textbf{15.86} & \cellcolor[HTML]{D9D9D9}0   & \cellcolor[HTML]{D9D9D9}\textbf{44} \\
TeMPO\_16$\times$16\_C         & 27  & 3.47     & \cellcolor[HTML]{D9D9D9}34.58     & \cellcolor[HTML]{D9D9D9}35   & \cellcolor[HTML]{D9D9D9}251  & 23 & 3.86  & \cellcolor[HTML]{D9D9D9}32.27 & \cellcolor[HTML]{D9D9D9}24  & \cellcolor[HTML]{D9D9D9}1571  & 15 & 5.48  & \cellcolor[HTML]{D9D9D9}\textbf{28.31} & \cellcolor[HTML]{D9D9D9}0   & \cellcolor[HTML]{D9D9D9}\textbf{169} \\
TeMPO\_32$\times$32\_C         & 49  & 5.91     & \cellcolor[HTML]{D9D9D9}62.57     & \cellcolor[HTML]{D9D9D9}113   & \cellcolor[HTML]{D9D9D9}1125  & 46 & 6.11  & \cellcolor[HTML]{D9D9D9}60.45 & \cellcolor[HTML]{D9D9D9}57  & \cellcolor[HTML]{D9D9D9}8057  & 31 & 10.96 & \cellcolor[HTML]{D9D9D9}\textbf{53.27} & \cellcolor[HTML]{D9D9D9}0   & \cellcolor[HTML]{D9D9D9}\textbf{425} \\
GWOR\_16$\times$16\_C          & 30 & 3.09  & \cellcolor[HTML]{D9D9D9}16.72 & \cellcolor[HTML]{D9D9D9}35  & \cellcolor[HTML]{D9D9D9}19    & 27 & 2.78  & \cellcolor[HTML]{D9D9D9}15.02 & \cellcolor[HTML]{D9D9D9}26  & \cellcolor[HTML]{D9D9D9}15    & 21 & 2.80  & \cellcolor[HTML]{D9D9D9}\textbf{12.03} & \cellcolor[HTML]{D9D9D9}0   & \cellcolor[HTML]{D9D9D9}\textbf{7} \\
GWOR\_32$\times$32\_C          & 62 & 7.22  & \cellcolor[HTML]{D9D9D9}35.26 & \cellcolor[HTML]{D9D9D9}126 & \cellcolor[HTML]{D9D9D9}60    & 56 & 6.81  & \cellcolor[HTML]{D9D9D9}31.26 & \cellcolor[HTML]{D9D9D9}44  & \cellcolor[HTML]{D9D9D9}45    & 44 & 7.82  & \cellcolor[HTML]{D9D9D9}\textbf{25.40} & \cellcolor[HTML]{D9D9D9}0   & \cellcolor[HTML]{D9D9D9}\textbf{11} \\
Benes\_16$\times$16\_C        & 33 & 4.42  & \cellcolor[HTML]{D9D9D9}22.69 & \cellcolor[HTML]{D9D9D9}76 & \cellcolor[HTML]{D9D9D9}721 & 26 & 4.28  & \cellcolor[HTML]{D9D9D9}18.84 & \cellcolor[HTML]{D9D9D9}3   & \cellcolor[HTML]{D9D9D9}175   & 23 & 3.57  & \cellcolor[HTML]{D9D9D9}\textbf{17.15} & \cellcolor[HTML]{D9D9D9}0   & \cellcolor[HTML]{D9D9D9}\textbf{33} \\
Benes\_32$\times$32\_C        & 56  & 9.24     & \cellcolor[HTML]{D9D9D9}36.76     & \cellcolor[HTML]{D9D9D9}365   & \cellcolor[HTML]{D9D9D9}5117  & 68 & 9.08  & \cellcolor[HTML]{D9D9D9}42.39 & \cellcolor[HTML]{D9D9D9}78  & \cellcolor[HTML]{D9D9D9}1690  & 58 & 6.51  & \cellcolor[HTML]{D9D9D9}\textbf{36.56} & \cellcolor[HTML]{D9D9D9}0   & \cellcolor[HTML]{D9D9D9}\textbf{251} \\
\hline
Geo-mean             & -  & 5.39     & \cellcolor[HTML]{D9D9D9}29.21     & \cellcolor[HTML]{D9D9D9}-   & \cellcolor[HTML]{D9D9D9}800   & -  & 5.27  & \cellcolor[HTML]{D9D9D9}27.05 & \cellcolor[HTML]{D9D9D9}-   & \cellcolor[HTML]{D9D9D9}1078  & -  & 5.61  & \cellcolor[HTML]{D9D9D9}\textbf{24.58} & \cellcolor[HTML]{D9D9D9}-   & \cellcolor[HTML]{D9D9D9}\textbf{155} \\
Ratio                & -  & 1     & \cellcolor[HTML]{D9D9D9}1     & \cellcolor[HTML]{D9D9D9}-   & \cellcolor[HTML]{D9D9D9}1     & -  & 0.97  & \cellcolor[HTML]{D9D9D9}0.92  & \cellcolor[HTML]{D9D9D9}-   & \cellcolor[HTML]{D9D9D9}1.33  & -  & 1.04  & \cellcolor[HTML]{D9D9D9}\textbf{0.84}  & \cellcolor[HTML]{D9D9D9}-   & \cellcolor[HTML]{D9D9D9}\textbf{0.19} \\
\hline
\end{tabular}
}
\label{tab:compact_comparison}
\vspace{-3pt}
\end{table*}

\begin{figure}[t]
    \centering
    \includegraphics[width=0.85\columnwidth]{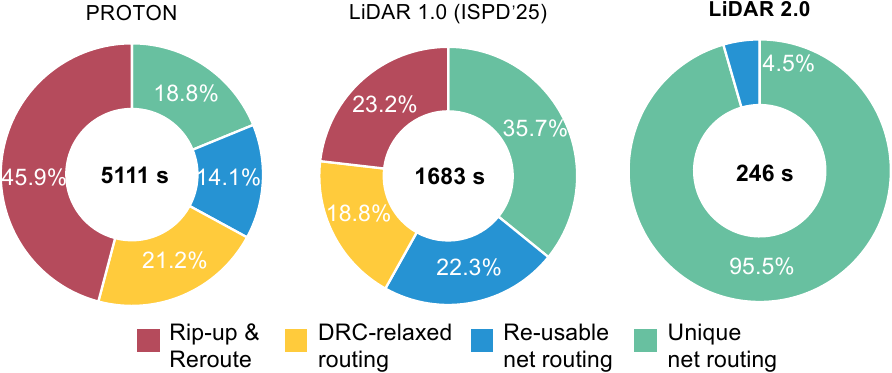}
    \caption{Runtime breakdown of PROTON~\cite{proton}, \name 1.0~\cite{PD_ISPD2025_Zhou}, and \name 2.0 in percentages on \texttt{Benes\_32$\times$32} with total runtime marked in the center.}
    \label{fig:runtime}
    \vspace{-14pt}
\end{figure}

We further evaluate routing on compact benchmarks with higher component density and more crossings. 
Only one rip-up and reroute round is applied, as limited space makes further iterations ineffective. 
Figure~\ref{fig:layout_compact} shows the final layouts by \name2.0.

As in Table~\ref{tab:compact_comparison}, \name2.0 delivers \textbf{nearly DRV-free results} with an average \textbf{16\% IL reduction}, \textbf{5.16$\times$ speedup} over \texttt{PROTON}, and \textbf{6.95$\times$ speedup} with \textbf{9\% lower IL} vs. \name1.0. 
On \texttt{TeMPO}, \texttt{PROTON} runs faster only because complete port blocking forces early termination.
For compact cases, \name1.0 shows more DRVs and runtime, especially on new benchmarks. 
In contrast, \name2.0’s hierarchical routing reuses subcircuits (\texttt{TeMPO}, \texttt{Benes}), greatly improving runtime. 
Intra-group ordering and proactive crossing-space reservation further reduce conflicts, enhance quality, and avoid redundant search. 
Thus, \name2.0 achieves DRV-free routing even on high-crossing \texttt{Benes}. 
For \texttt{GWOR}, lacking group data, \name2.0 still succeeds by dynamically refining order and port regions during reroute.

Figure~\ref{fig:runtime} shows the runtime breakdown on \texttt{Benes}\_32$\times$32. 
The speedup of \name2.0 comes from reducing conflicts, eliminating redundant search and RR iterations, and hierarchically reusing results.

\vspace{-8pt}
\subsection{Results on Different Crossing Sizes and Bend Radii}
\begin{table*}[t]
\centering
\caption{\name2.0 can handle various sizes of waveguide crossings in PICs with different layout densities.}
\renewcommand{\arraystretch}{1.2}
\small
\resizebox{\textwidth}{!}{
\begin{tabular}{cccccccccccccccc}
\hline \addlinespace[0.4ex]
\multirow{2}{*}{\raisebox{-0.5ex}[0pt]{Benchmarks}} & \multicolumn{5}{c}{Crossing size = $10 \times 10 \mu m^2$} & \multicolumn{5}{c}{Crossing size = $15 \times 15 \mu m^2$} & \multicolumn{5}{c}{Crossing size = $20 \times 20 \mu m^2$} \\
[-0.1em]
\cmidrule(lr){2-6} \cmidrule(lr){7-11} \cmidrule(lr){12-16}\addlinespace[-0.3ex]
& \#CR & WL & $IL_{\max}\downarrow$ & \#DRV $\downarrow$ & Time $\downarrow$(s) 
& \#CR & WL & $IL_{\max}\downarrow$ & \#DRV $\downarrow$ & Time $\downarrow$(s) 
& \#CR & WL & $IL_{\max}\downarrow$ & \#DRV $\downarrow$ & Time $\downarrow$(s) \\
\hline
ADEPT\_16$\times$16\_C    & 21 & 8.35 & 21.02 & 0 & 236 & 21 & 8.54 & 21.08 & 0 & 299 & 21 & 9.08 & 20.96 & 3 & 379 \\
ADEPT\_16$\times$16\_L    & 21 & 8.84 & 21.05 & 0 & 339 & 21 & 8.80 & 21.06 & 0 & 347 & 21 & 9.18 & 21.13 & 1 & 373 \\
TeMPO\_16$\times$16\_C    & 15 & 5.48 & 28.31 & 0 & 119 & 15 & 5.41 & 28.30 & 0 & 127 & 15 & 5.33 & 28.29 & 0 & 131 \\
TeMPO\_16$\times$16\_L    & 15 & 4.86 & 28.22 & 0 & 132 & 15 & 4.79 & 28.21 & 0 & 136 & 15 & 4.71 & 28.19 & 0 & 139 \\
Benes\_16$\times$16\_C   & 23 & 3.57 & 17.15 & 0 & 33  & 23 & 3.45 & 17.13 & 0 & 34  & 23 & 3.34 & 17.15 & 0 & 34  \\
Benes\_16$\times$16\_L   & 22 & 3.70 & 16.68 & 0 & 34  & 22 & 3.57 & 16.62 & 0 & 28  & 23 & 3.49 & 17.23 & 0 & 35  \\
\hline
\end{tabular}
}
\vspace{-5pt}
\label{tab:crossing_size_comparison}
\end{table*}

\begin{table*}[t]
\centering
\caption{\name2.0 can flexibly support PIC routing under various user-specified bend radii $r$ ($\mu m$). 
Different radii are shown according to the chip area and layout compactness of each benchmark.}
\renewcommand{\arraystretch}{1.1}
\small
\resizebox{0.95\textwidth}{!}{
\begin{tabular}{cccccccccccccc}
\hline \addlinespace[0.4ex]
\multirow{2}{*}{\raisebox{-0.5ex}[0pt]{Metrics}}
& \multicolumn{5}{c}{GWOR\_32$\times$32} 
& \multicolumn{2}{c}{Clements\_16$\times$16\_C} 
& \multicolumn{2}{c}{ADEPT\_16$\times$16\_C} 
& \multicolumn{2}{c}{Benes\_16$\times$16\_C} 
& \multicolumn{2}{c}{TeMPO\_16$\times$16\_C} \\
[-0.1em]
\cmidrule(lr){2-6} \cmidrule(lr){7-8} \cmidrule(lr){9-10} \cmidrule(lr){11-12} \cmidrule(lr){13-14} \addlinespace[-0.3ex]
& $r$=10 & $r$=20 & $r$=40 & $r$=80 & $r$=100 
& $r$=10 & $r$=15 
& $r$=10 & $r$=15 
& $r$=5 & $r$=10 
& $r$=5 & $r$=10 \\
\hline
WL     & 7.77 & 7.77 & 7.79 & 7.85 & 8.06 & 3.49 & 3.33 & 8.55 & 8.52 & 3.57 & 3.76 & 5.48 & 5.41 \\
\#CR     & 44   & 44   & 44   & 44   & 44   & 0    & 3    & 21   & 22   & 23   & 23   & 15    & 15   \\
$IL_{\max}$ & 26.40 & 25.40 & 25.40 & 25.41 & 25.44 & 26.54 & 26.25 & 21.06 & 21.62 & 17.15 & 17.32 & 28.32 & 27.90 \\
\#DRV    & 0    & 0    & 0    & 0    & 3    & 0    & 3    & 0    & 2    & 0    & 0    & 0    & 0    \\
Time (s) & 10   & 10   & 10   & 14   & 16   & 96   & 111  & 298  & 339  & 33   & 33   & 169   & 172  \\
\hline
\end{tabular}
}
\label{tab:bend_radius_comparison}
\end{table*}

Different PIC technologies impose varying constraints on crossing sizes and bending radii.
Table~\ref{tab:crossing_size_comparison} and Table~\ref{tab:bend_radius_comparison} show the routing results of \name2.0 under different crossing sizes and bend radii. We synthetically stretch the device only to increase the routing difficulty and test our router's capability, regardless of its actual IL value. Benchmarks with an \texttt{L} suffix indicate larger component pitch.
As expected, larger crossing sizes and bend radii increase area usage and routing difficulty.
Still, \name2.0 consistently generates DRV-free layouts across compact and spacious benchmarks, with only minor DRVs in \texttt{ADEPT} cases.
\name2.0 supports the insertion of various crossing sizes and remains robust under different bend radii.
It successfully routes \texttt{GWOR\_32$\times$32} even with large bend radii up to 80~$\mu m$, which is typical in silicon nitride platforms.
For silicon photonics, where the bending radius is typically 5–10~$\mu m$, \name2.0 performs reliably.
 Feasible solutions remain available at 15~$\mu m$ despite minor DRVs, except for the \texttt{TeMPO} and \texttt{Benes} benchmarks, where no valid routing solution exists due to insufficient spacing.

\vspace{-8pt}
\subsection{Discussion}
\label{sec:Ablation}
\begin{figure}
    \centering
    \includegraphics[width=0.95\columnwidth]{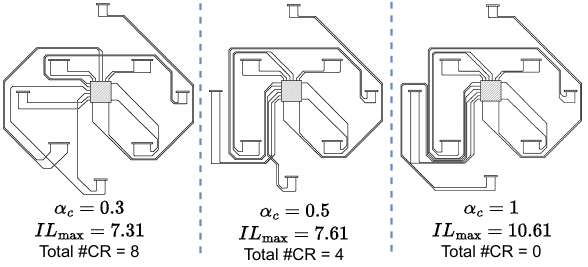}
    \vspace{-5pt}
    \caption{Layout of \texttt{Light}\_a of different crossing loss.
    }
    \label{fig:NOC}
    \vspace{-5pt}
\end{figure}

\begin{table}
\centering
\vspace{-5pt}
\caption{Compare to manual routing on small PIC circuits.}
\vspace{-5pt}
\resizebox{\columnwidth}{!}{
\begin{tabular}{ccccc}
\hline \addlinespace[0.4ex]
\multirow{2}{*}{\raisebox{-0.5ex}[0pt]{Metrics}} & \multicolumn{2}{c}{Light\_a}    & \multicolumn{2}{c}{ADEPT\_4$\times$4}  \\  
[-0.1em]
\cmidrule(lr){2-3} \cmidrule(lr){4-5} \addlinespace[-0.3ex]
                           & Manual        & \textbf{\name 2.0}           & Manual        & \textbf{\name 2.0} \\       \hline
\#CR   &  0        & 0     & 10                           & 10    \\
WL (mm)   & 31.89    & \textbf{29.99} &  \textbf{2.30}   & 2.33  \\
$IL_{max}\downarrow$   & 7.85     & \textbf{7.61}  & \textbf{12.95} & 13.14 \\
DRV     & 0      & 0       & 0      & 0      \\
Time   & $\sim$2h & \textbf{103s}  & $\sim$2h    & \textbf{26s} \\  
\hline
\end{tabular}
}
\vspace{-5pt}
\label{tab:manual}
\end{table}

\begin{table}
\centering
\caption{Congestion penalty with different crossing costs.}
\vspace{-5pt}
\renewcommand{\arraystretch}{1.1}
\resizebox{\columnwidth}{!}{
\begin{tabular}{ccccc}
\hline \addlinespace[0.4ex]
\multirow{2}{*}{\raisebox{-0.5ex}[0pt]{Metrics}} & \multicolumn{2}{c}{High Crossing Cost $\alpha_c=1$}    & \multicolumn{2}{c}{Low Crossing Cost $\alpha_c=0.3$}   \\  
[-0.1em]
\cmidrule(lr){2-3} \cmidrule(lr){4-5} \addlinespace[-0.3ex]
                           & w/o GCP        & \textbf{\name 2.0}           & w/o GCP        & \textbf{\name 2.0} \\       \hline
\#CR     & 6     & 0     & 5      & 5     \\
WL (mm)   & 20.72    & 29.99     & 25.11  & 26.04  \\
$IL_{max}\downarrow$   & 15.21    & 10.61    & 7.18   & 7.31  \\
DRV     & 0      & 0       & 1      & 0      \\
Time (s)    & 129      & 103      & 261     & 197  \\ 
\hline
\end{tabular}
}
\vspace{-10pt}
\label{tab:ablation}
\end{table}

\begin{table}[t]
\centering
\vspace{-5pt}
\caption{Ablation results on representative benchmarks. 
Removing staggered port offset, group-based net ordering, or crossing space leads to more \#DRV.}
\renewcommand{\arraystretch}{1.3}
\small
\resizebox{\columnwidth}{!}{
\begin{tabular}{ccccccc}
\hline \addlinespace[0.4ex]
\multirow{2}{*}{\raisebox{-0.5ex}[0pt]{Benchmarks}}
& \multicolumn{2}{c}{w/o Staggered offset} & \multicolumn{2}{c}{w/o crossing space} & \multicolumn{2}{c}{w/o group-based net order} \\
[-0.1em]
\cmidrule(lr){2-3} \cmidrule(lr){4-5} \cmidrule(lr){6-7} \addlinespace[-0.3ex]
& $IL_{\max}$ & \#DRV  
& $IL_{\max}$ & \#DRV 
& $IL_{\max}$ & \#DRV \\
\hline
ADEPT\_16$\times$16\_C    & 18.95 & 37 & 19.97 & 6  & 21.63 & 4  \\
ADEPT\_16$\times$16\_L    & 19.13 & 33 & 20.34 & 3  & 21.95 & 3  \\
TeMPO\_16$\times$16\_C    & 28.49 & 27 & 28.31 & 0  & 31.87 & 16 \\
TeMPO\_16$\times$16\_L    & 28.72 & 26 & 28.22 & 0  & 32.64 & 16 \\
Benes\_16$\times$16\_C    & 14.62 & 21 & 16.24 & 32 & 16.83 & 2  \\
Benes\_16$\times$16\_L    & 14.37 & 21 & 16.93 & 30 & 17.21 & 4 \\
\hline
\end{tabular}
}
\vspace{-5pt}
\label{tab:ablation_offset}
\end{table}

\begin{table}[t]
\centering
\caption{Proposed offset neighbors reduce \#DRV, $IL_{max}$, and runtime on TeMPO.
\emph{C+} means more compact layout where x/y pitch drops from 25/50 to 20/15 $\mu m$.}
\renewcommand{\arraystretch}{1.2}
\small
\resizebox{\columnwidth}{!}{
\begin{tabular}{ccccccc}
\hline \addlinespace[0.4ex]
\multirow{2}{*}{\raisebox{-0.5ex}[0pt]{Benchmarks}}
& \multicolumn{3}{c}{w/o Offset Neighbor} 
& \multicolumn{3}{c}{w/ Offset Neighbor} \\
[-0.1em]
\cmidrule(lr){2-4} \cmidrule(lr){5-7} \addlinespace[-0.3ex]
& $IL_{\max}$ & \#DRV & Time (s) 
& $IL_{\max}$ & \#DRV & Time (s) \\
\hline
TeMPO\_8$\times$8\_C+    & 18.00  & 22  & 96   & 15.86  & 0 & 44   \\
TeMPO\_8$\times$8\_C    & 16.02  & 0   & 67    & 15.14  & 0 & 40   \\
TeMPO\_16$\times$16\_C+  & 34.12  & 71 & 470   & 28.03  & 0 & 144  \\
TeMPO\_16$\times$16\_C  & 28.58  & 0   & 215   & 28.31  & 0 & 119  \\
TeMPO\_32$\times$32\_C+  & 58.57  & 157  & 1713  & 52.28  & 0 & 494  \\
TeMPO\_32$\times$32\_C & 53.94  & 0   & 673   & 53.27  & 0 & 425  \\
\hline
\end{tabular}
}
\vspace{-10pt}
\label{tab:offset_effect}
\end{table}

\noindent\underline{\textbf{Manual Routing Result}}.~ 
We conducted a manual routing comparison on small circuits, \texttt{Light\_a} and \texttt{ADEPT\_4$\times$4}, as shown in TABLE~\ref{tab:manual}. 
It can be observed that the routing quality of \name~2.0 is comparable to that of manual routing, and even achieves lower insertion loss on \texttt{Light\_a}. 
However, manual routing required approximately two hours to complete for these small circuits, and as the circuit scale increases, manual efforts become impractical.

\noindent\underline{\textbf{Spacing Reservation}}.~
We propose several techniques for reserving spacing, including \emph{staggered offset port region}, \emph{reserved crossing spacing}, \emph{group congestion penalty}, and \emph{group-based net ordering} to mitigate routing conflicts in multiport components.
TABLE~\ref{tab:ablation_offset} highlights their impact on representative circuits. 
It can be observed that the \emph{staggered offset port region} is most critical: removing it causes many routing failures. 
Without the \emph{crossing space penalty}, the \texttt{Benes} circuit suffers heavily due to consecutive crossings, while \texttt{TEMPO} is less affected since its crossings are independent. 
Disabling \emph{group-based net ordering}, however, triggers numerous DRVs in \texttt{TEMPO} because of high-fanout components.

We also assess the proposed GCP in guiding crossing optimization under different crossing weights $\alpha_c$ (Table~\ref{tab:ablation}).
With larger $\alpha_c$, the router avoids crossings to reduce maximum insertion loss ($IL_{\text{max}}$); with smaller $\alpha_c$, it favors shorter paths even with extra crossings, further improving $IL_{\text{max}}$.

Without GCP, more crossings arise as $\alpha_c$ increases, since earlier nets congest low-crossing paths. 
With GCP, \textbf{routing conflicts drop notably}, allowing efficient paths with fewer crossings and lower IL. 
In Fig.~\ref{fig:NOC}, $\alpha_c$ flexibly balances waveguide length and crossings to meet performance needs.

\noindent\underline{\textbf{Crossing-Waveguide Optimization}}.~
As shown in Table~\ref{tab:stat} (\textbf{PROTON}~\cite{proton} vs. \name2.0), our proposed crossing optimization strategy introduces an extra runtime penalty on the \texttt{Light} benchmark, but it leads to higher solution quality as it reduces the crossings between IOs and the switch, given the high crossing penalty.

\noindent\underline{\textbf{$A^\star$ Offset Neighbor}}.~
Offset neighbors help eliminate unnecessary bending detours caused by port misalignment, thereby reducing routing conflicts and search overhead.
We evaluate them on a compact \texttt{TeMPO\_C+} variant with x/y pitch reduced from 25/50~$\mu m$ to 20/15~$\mu m$ (Table~\ref{tab:offset_effect}).
In such constrained layouts, routing without offset neighbors often causes rule violations.
Even in spacious cases, enabling them yields consistently faster runtime.

\section{Conclusion}
\label{sec:Conclusion}
In this work, we present \name, an open-source detailed router for PICs. 
It features a non-Manhattan curvy-aware A$^\ast$ engine with enhanced port assignment, adaptive crossing insertion, congestion-aware group-based net ordering, and crossing optimization to address unique PIC constraints while minimizing critical path insertion loss. 
We further extend it into a scalable, conflict-resilient framework, \name2.0, by adding hierarchical routing reuse, intra-group ordering, offset-neighbor search, and crossing-space preservation. 
These enhancements markedly improve routability under tight layouts and high crossing density. 
\name2.0 achieves nearly DRV-free layouts on challenging benchmarks, with up to \textbf{9\% lower IL} and \textbf{6.95$\times$ speedup} over \name1.0, advancing automation for large-scale PIC design.

\vspace{-.05in}

\begin{IEEEbiography}[{\includegraphics[width=1in,height=1.24in,clip,keepaspectratio]{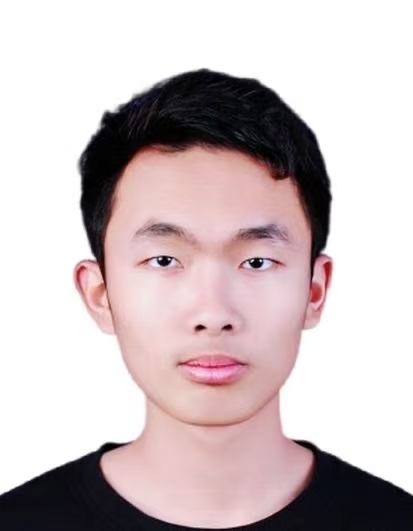}}]{Hongjian Zhou}
received the B.E.~degree in electronic information engineering from Central South University, Changsha, China, in 2021, and the M.S. degree in electronic science and technology from ShanghaiTech University, Shanghai, China, in 2024.  He is currently pursuing the Ph.D. degree with the School of Electrical, Computer and Energy
Engineering, Arizona State University, AZ, USA, under the supervision of Prof. Jiaqi Gu. His interests include physical design, electronic-photonic design automation, and machine learning for EDA. He has received the Best Paper Award Nomination at ASP-DAC 2024.
\end{IEEEbiography}
\vspace{-0.2in}

\begin{IEEEbiography}[{\includegraphics[width=1in,height=1.24in,clip,keepaspectratio]{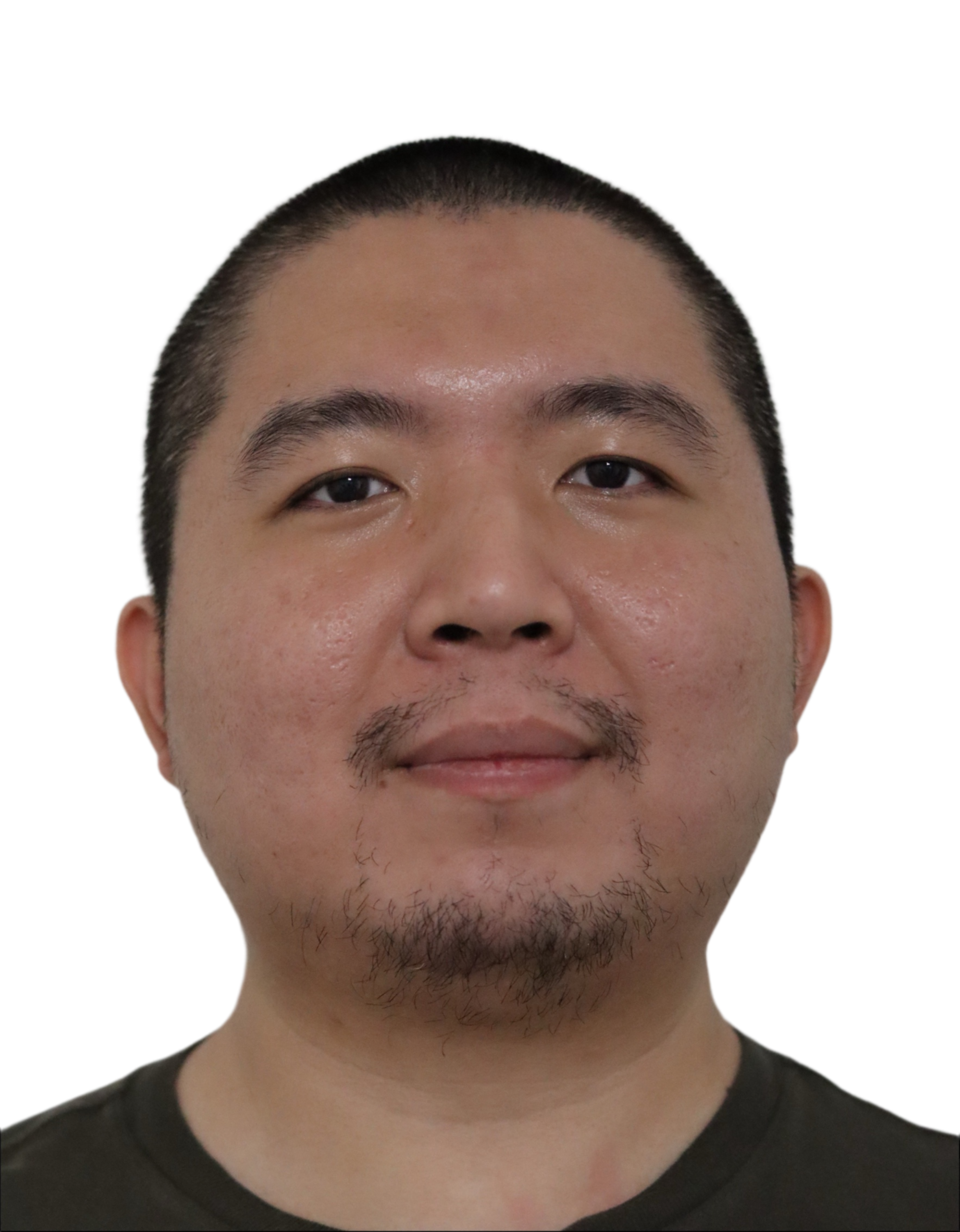}}]{Haoyu Yang} received the Ph.D. degree from The Chinese University of Hong Kong, Hong Kong, in 2020. He is currently a Senior Research Scientist with the Design Automation Research Group, NVIDIA, Austin, TX, USA. His research interests include AI for electronic design automation, AI for computational lithography, GPU acceleration, and LLM-Aided Design. Dr. Yang received the Best Paper Award from IEEE TSM in 2022 and the Best Paper Award Nomination from ASPDAC 2019. He is also the recipient of the 2019 Nick Cobb Scholarship from SPIE and Mentor Graphics.
\end{IEEEbiography}
\vspace{-0.2in}

\begin{IEEEbiography}[{\includegraphics[width=1in,height=1.24in,clip,keepaspectratio]{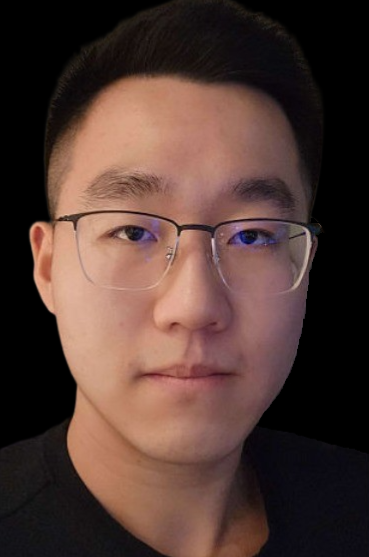}}]{Ziang Yin} received the B.S. degree in Electrical and Computer Engineering from the University of Washington, Seattle, WA, USA, in 2022 and the M.S. degree in Electrical and Computer Engineering from the same institution in 2023. He is currently pursuing the Ph.D. degree in the School of Electrical, Computer, and Energy Engineering at Arizona State University, Tempe, AZ, USA, where he works as a Graduate Research Assistant in Prof. Jiaqi Gu’s Scope X Lab. His research interests span efficient machine learning and intelligent computing, including electronic-photonic accelerator architecture, cross-layer circuit-architecture-algorithm co-design, and AI/CAD techniques for optical and quantum systems. He has authored or co-authored more than ten peer-reviewed journal and conference papers—appearing in venues such as Science Robotics, Journal of Applied Physics, ACM/IEEE DAC, and ICCAD. He is a student member of IEEE-HKN.
\end{IEEEbiography}
\vspace{-0.2in}

\begin{IEEEbiography}[{\includegraphics[width=1in,height=1.26in,clip,keepaspectratio]{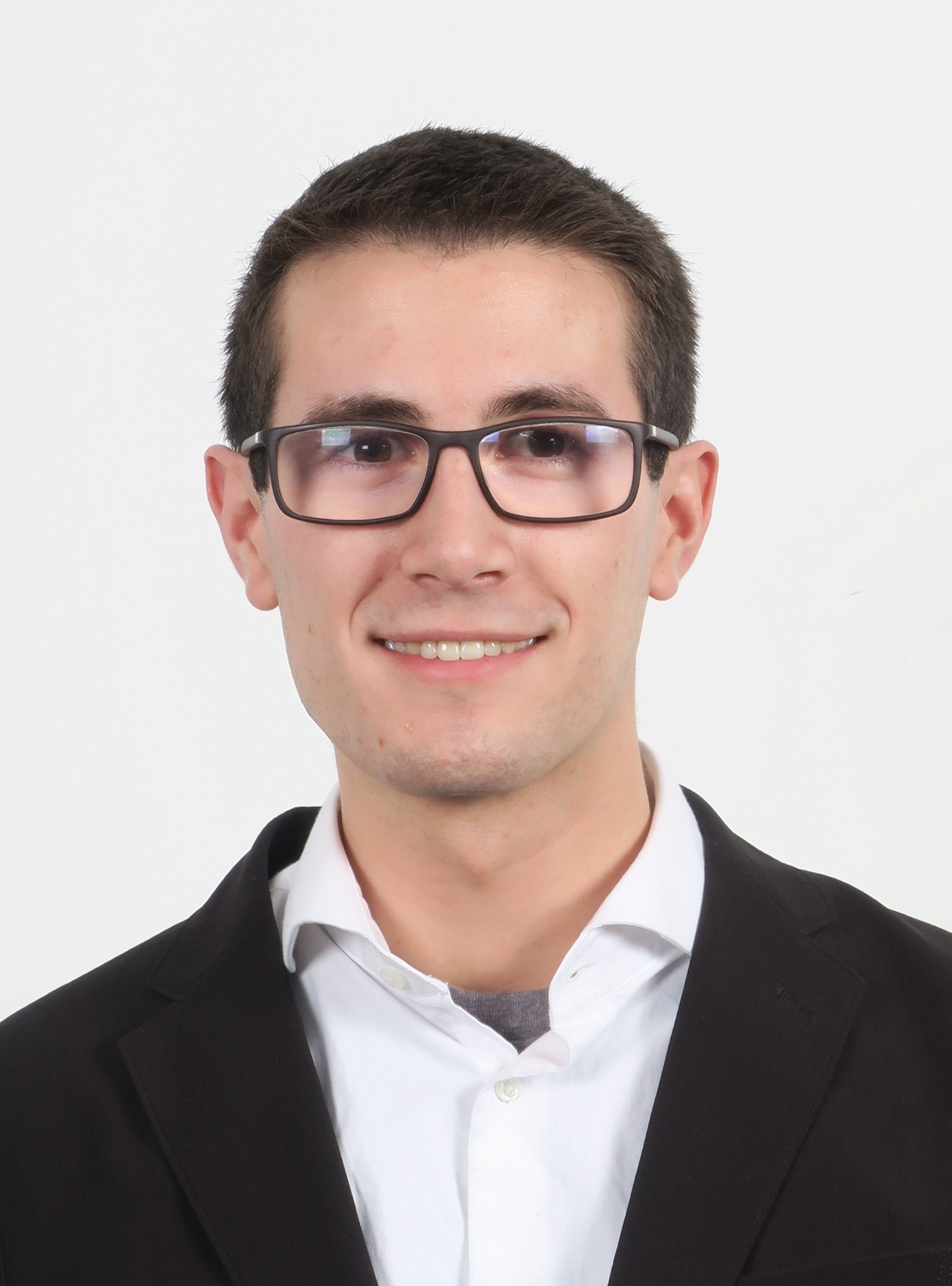}}]{Nicholas Gangi} received the B.S. degree in Biomedical Engineering from Rensselaer Polytechnic Institute, Troy, NY in 2022.  He is continuing his studies at Rensselaer and is currently pursuing the M.S. and Ph.D. degrees with the School of Electrical, Computer and Systems Engineering, under the supervision of Prof. Zhaoran (Rena) Huang.  His interests include the design and fabrication of photonic devices and architectures relevant to photonic computing, machine learning, and communication.
\end{IEEEbiography}
\vspace{-0.2in}

\begin{IEEEbiography}[{\includegraphics[width=1in,height=1.24in,clip,keepaspectratio]{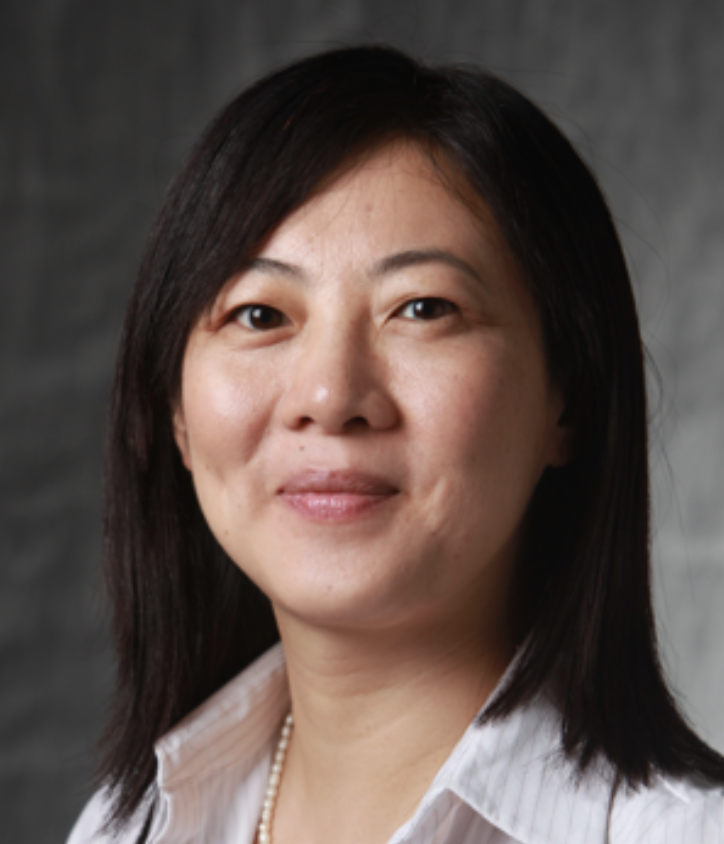}}]{Zhaoran (Rena) Huang} (Senior Member, IEEE) received her B.S. degree from the Beijing Institute of Technology, Beijing, China, and her M.Sc. and Ph.D. degrees from the Georgia Institute of Technology, Atlanta, GA, USA. She is currently an Associate Professor in the Department of Electrical, Computer, and Systems Engineering at Rensselaer Polytechnic Institute, Troy, NY, USA. Her research interests include SiGe modulators, high-speed Schottky photodiodes, integrated photonics, silicon photonics, optical reservoir computing, and photonic quantum information processing. Prof. Huang is a member of OSA and SPIE. She has served as a guest editor for the IEEE Journal of Lightwave Technology and the Journal of Applied Physics, and currently serves as an Associate Editor for the IEEE Photonics Technology Letters.
\end{IEEEbiography}
\vspace{-0.2in}

\begin{IEEEbiography}[{\includegraphics[width=1in,height=1.24in,clip,keepaspectratio]{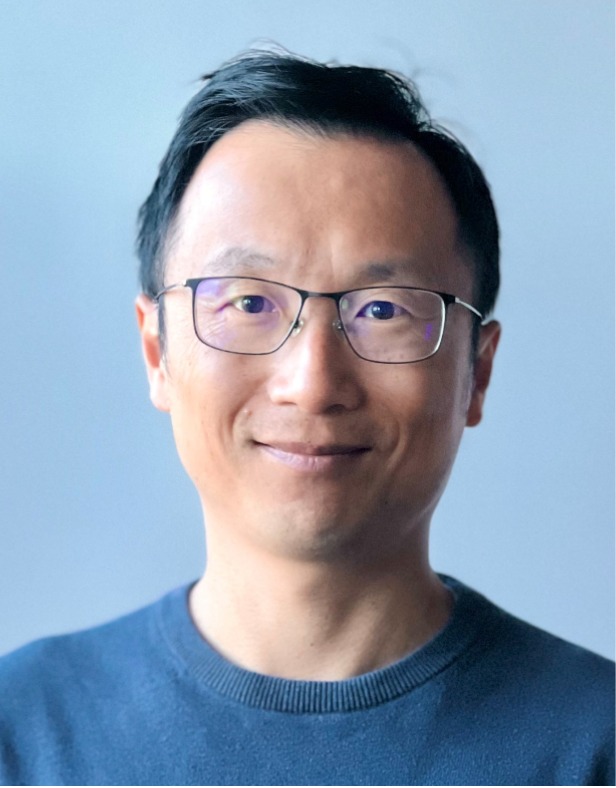}}]{Haoxing Ren}  (S’99-M’00-SM’09-F’24) received the B.S/M.S. degrees in electrical engineering from Shanghai Jiao Tong University, Shanghai, China, the M.S. degree in computer engineering from Rensselaer Polytechnic Institute, Troy, NY, USA, and the Ph.D. degree in computer engineering from the University of Texas at Austin, Austin, TX, USA. He is currently the Director of Design Automation Research at NVIDIA, focusing on leveraging machine learning and GPU-accelerated tools to enhance chip design quality and productivity. He has over 25 years of industrial EDA research and development experience at IBM and NVIDIA. He holds over thirty patents and has co-authored over 100 papers and books, including a book on Machine Learning for EDA and several book chapters in EDA. He received several prestigious awards for his work, including the IBM Corporate Award and Best Paper Awards at ISPD, DAC, TCAD, MLCAD and IEEE LAD. He serves in the organization and steering committees of international conferences such as ICCAD and ISPD and as the conference chair at ICLAD.
\end{IEEEbiography}
\vspace{-0.2in}

\begin{IEEEbiography}[{\includegraphics[width=1in,height=1.24in,clip,keepaspectratio]{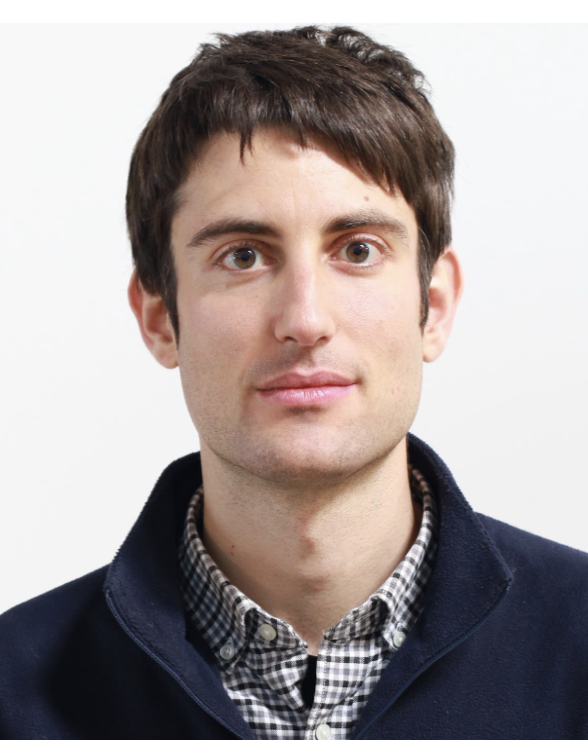}}]{Joaquin Matres} received the Engineering degree in Telecommunications in 2009, the M.Sc. degree in 2010, and the Ph.D. degree in 2014 from the Universidad Politécnica de Valencia and the Nanophotonics Technology Center, Spain. His doctoral research focused on the nonlinear properties of silicon-based nanophotonic waveguides, leading to the development of CMOS-compatible all-optical switches and logic gates. This work involved in-house electron-beam lithography and collaborations with IMEC and CEA-LETI.
He conducted part of his research at the Optoelectronics Research Centre at the University of Southampton, U.K., and at the University of California, Davis, where he collaborated with industry partners to develop wavelength-selective filters. 
He later joined Intel Corporation as an intern and subsequently worked at Hewlett Packard Labs, contributing to the development of hybrid tunable lasers, wavelength-selective switches, and microring-based optical transceivers.
Driven by a passion for applying Python to chip design, he founded the open-source project GDSFactory in 2019. Aimed at modernizing and democratizing chip design, GDSFactory has become a widely adopted tool with over 2 million downloads.
He is currently leading the development of GDSFactory+, which provides enterprise-grade support for GDSFactory, including access to foundry PDKs, schematic capture, simulation, and verification tools (DRC, LVS). He actively collaborates with academic institutions such as the University of Toronto, the Max Planck Institute, and Arizona State University to push the boundaries of next-generation chip design.
\end{IEEEbiography}
\vspace{-0.2in}

\begin{IEEEbiography}[{\includegraphics[width=1in,height=1.24in,clip,keepaspectratio]{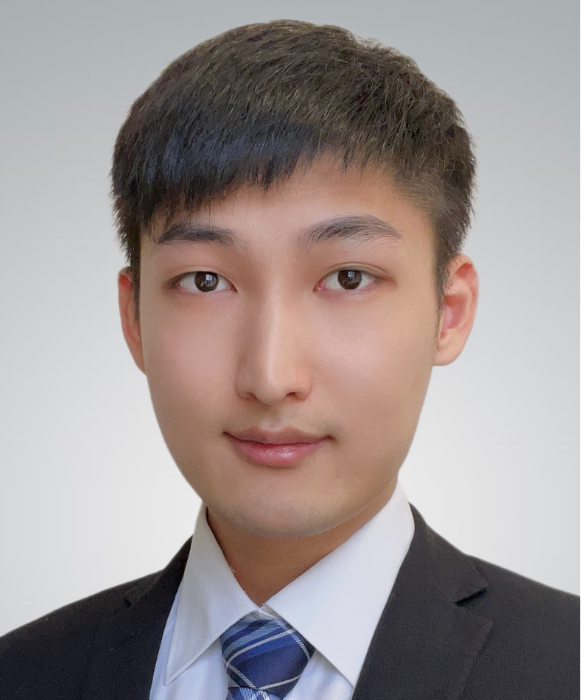}}]{Jiaqi Gu} (S'19 - M'23)
received the B.E.~degree in Microelectronic Science and Engineering from Fudan University, Shanghai, China in 2018. 
He is currently a post-graduate student studying for his Ph.D. degree in the Department of Electrical and Computer Engineering, The University of Texas at Austin, Austin, TX, USA. 
He is an Assistant Professor in the School of Electrical, Computer and Energy Engineering at Arizona State University (ASU), Tempe, AZ, USA. 
His current research interests include machine learning, efficient algorithms and architecture design for high-performance AI, next-generation AI computing with emerging technology, and electronic-photonic design automation.
He has published over 90 journal articles and refereed conference papers in the related field.
He has received the Best Paper Award at ASP-DAC 2020, the Best Paper Finalist at DAC 2020, the Best Poster Award at NSF Workshop on Machine Learning Hardware (2020), and the ACM Student Research Competition Grand Finals First Place (2021), the Best Paper Award at IEEE TCAD 2021, won the Robert S. Hilbert Memorial Optical Design Competition 2022, Graduate School Outstanding Dissertation Award at UT Austin (2024), the Best Paper Award Candidate at DATE 2025, and the NVIDIA NVIDIA Academic Grant Program Award (2025).
\end{IEEEbiography}
\vspace{-0.2in}

\end{document}